# TOWARDS GEOSTRATEGIC CRITICAL MINERALS AND MATERIALS RESILIENCE

## SECURE SUPPLY-CHAIN AND CRITICALITY ANALYSES FOR QUANTUM TECHNOLOGIES IN ARCTIC AND SPACE ENVIRONMENTS

*Min-Ha Lee, Alan J. Hurd, Jolante Wieke Van Wijk, and Mauritz Kop*

**Authors and Affiliations:**

- **Min-Ha Lee**
  Centre for International Security and Cooperation (CISAC), Stanford University, Stanford, CA 94305, United States.
  Korea Institute of Industrial Technology (KITECH), Incheon 21999, Republic of Korea.
- **Alan J. Hurd**
  Division of Earth and Environmental Sciences, Los Alamos National Laboratory, PO Box 1663, Los Alamos, NM, USA 87545.
- **Jolante Wieke Van Wijk**
  Division of Earth and Environmental Sciences, Los Alamos National Laboratory, PO Box 1663, Los Alamos, NM, USA 87545.
- **Mauritz Kop**
  Centre for International Governance Innovation (CIGI), Waterloo, ON N2L 6R2, Canada.

**Corresponding Author:**

Mauritz Kop
Email: advies@airecht.nl

## Abstract

*This manuscript maps secure-supply and criticality risks for quantum technologies deployed in extreme environments, linking upstream critical minerals and materials (CMMs) to downstream system performance, continuity of security, and mission assurance. It adopts a reproducible "Critical Level I" screening method to identify materials whose supply concentration, essentiality, and limited mitigatability can create bottlenecks for quantum deployment. The analysis is structured around two use cases: (i) niobium as a key input for superconducting quantum computing and related manufacturing and toolchain dependencies; and (ii) space-qualified superconducting nanowire single-photon detectors (SNSPDs), alongside adjacent single-photon detector platforms such as SPADs, where radiation, thermal cycling, vibration, and electromagnetic interference can degrade device metrics and, in communications settings, threaten continuity of security. The manuscript further situates these dependencies within U.S.–China strategic competition over critical materials, refining capacity, export controls, and overseas mineral acquisitions, while also connecting them to standards-first governance, post-quantum cryptography migration, and the emerging security logic of quantum networking. It*

*argues that static national critical-minerals lists are insufficient for mission-relevant quantum technology and proposes a dedicated Quantum Criticality and Critical Minerals (QCCM) dashboard as a living decision-support tool for tracking concentration, substitutability, qualification bottlenecks, stockpiling gaps, and geopolitical stress signals across quantum platforms. The paper concludes with implications for substitution, diversification, stockpiling, shielding, qualification-by-design, and standards-aligned governance to support secure, sustained, and mission-relevant quantum deployment.*

*The Introduction frames the CMM supply-chain challenge in quantum technologies. Section 1 examines upstream CMM dependencies in quantum technology, Section 2 addresses stockpiling strategies, and Section 3 develops Use Case 1 through niobium for superconducting quantum computing and nickel for magnetic shielding. Section 4 develops Use Case 2 on quantum technology in harsh environments, particularly space-qualified SNSPDs, with Section 4.1 focusing on shielding and radiation-hardening materials. Section 5 translates these dependencies into security and deployment implications, including quantum-secure communications, mission assurance, and quantum–AI hybrids. Section 6 concludes with mitigation and research priorities.*

## Introduction

There have been significant developments across the quantum technology (QT) landscape over the past three years, especially in the private sector. At the end of 2023, Atom Computing announced the first quantum processor to exceed 1,000 qubits.[1] Soon after, IBM announced Heron, a quantum processor with significantly improved error rates, and Condor, a quantum processor with more than 1,000 qubits. In December 2024, Google unveiled its quantum computing (QC) chip "Willow," reporting a milestone in quantum error correction on a superconducting-qubit platform on the path toward fault-tolerant operation. In February 2025, Microsoft announced Majorana 1, a prototype processor powered by topological qubits (a "Topological Core" in which quantum states are protected by topology rather than low temperature). Amazon joined the chorus of announcements later that month with the company's first-generation quantum chip, 'Ocelot,' notable for its use of bosonic quantum error correction. Given the high rate of change in QT hardware, any account of recent progress will rapidly become dated.

Although the technologies behind these developments are different—a hallmark of the quantum innovation ecosystem, where the optimal pathway to realizing the full power of QC remains unresolved—they all depend on Critical Minerals and Materials (CMMs) as necessary inputs. For example, Atom Computing's technology uses strontium (Sr) and ytterbium (Yb), while IBM's processors reportedly use niobium (Nb), among other critical minerals. Google's processor uses indium (In) for the bump bonds between wafers, and Microsoft's processor uses indium arsenide–aluminum and indium-phosphate. The high-quality oscillators in Amazon's Ocelot are made from a thin film of tantalum (Ta). QC's reliance on CMMs—especially rare earths—extends beyond processors to a range of other components, such as dilution refrigerators, lasers, optics, and electromagnetic components.

[1] Alex Wilkins, Record-breaking quantum computer has more than 1000 qubits, New Scientist, 24 Oct. 2024, https://www.newscientist.com/article/2399246-record-breaking-quantum-computer-has-more-than-1000-qubits/.

In the public sector, as Vidal Bustamante and Burke argue in a March 2026 CNAS report on U.S. quantum supply chains, federal funding under the National Quantum Initiative has grown substantially since 2018 yet remains concentrated on fundamental research, with "less than 12 percent directed toward enabling technologies and manufacturing infrastructure."[2] Their analysis ties the strategic payoff of U.S. quantum leadership to industrial capacity: durable advantage requires the ability to manufacture quantum systems at scale, and persistent weaknesses in domestic supply chains, combined with reliance on suppliers in China and Russia, threaten to slow U.S. progress and deepen foreign dependence. On this view, a stronger enabling-technology base is the condition under which scientific achievement converts into economic and security gains.

Crucially, the same CMM-dependent enabling stack underpins the wider suite of QT beyond computing—including quantum sensing (QS) and timing, quantum optics, quantum networking and cryptography, and quantum simulation—so supply risk and mitigatability must be evaluated at the level of mission-relevant systems rather than isolated devices. These shared CMM dependencies, in turn, connect QT performance and deployment to geostrategic resilience: concentrated mining and refining capacity, trade frictions, and export-control dynamics can rapidly translate commercial dependence into strategic vulnerability, shaping which QT capabilities can be scaled, fielded, and sustained in Arctic and space environments, where qualification barriers and operational fragility amplify supply shocks. Accordingly, it is essential to identify technological and geopolitical mitigation pathways that reduce the mismatch between QT systems whose performance is governed by thermodynamic constraints and CMM supply chains whose resilience is constrained by extraction, processing, and delivery kinetics.

As QT moves from theory to deployment, the governance question becomes inseparable from the materials question. Recent scholarship argues that early, globally harmonized standards for terminology, benchmarking, safety, interoperability, and quality management can reduce fragmentation before hard law freezes still-immature architectures,[3] while responsible-quantum-innovation frameworks emphasize that safeguarding, stakeholder engagement, and long-horizon legitimacy must be built into the design cycle early rather than retrofitted later.[4] Supply-chain resilience is accordingly a constitutive element of responsible quantum technology (RQT) design: concentrated, brittle, or strategically exploitable materials dependencies undermine the long-horizon legitimacy and mission trustworthiness that responsible innovation frameworks require.

---

[2] C. M. Vidal Bustamante and J. Burke, Quantum's Industrial Moment: Strengthening U.S. Quantum Supply Chains for Scalable Advantage, Center for a New American Security (March 2026), https://www.cnas.org/publications/reports/quantums-industrial-moment, at 1-2.

[3] Mateo Aboy, Urs Gasser, I. Glenn Cohen & Mauritz Kop, *Quantum Technology Governance: A Standards-First Approach*, 389 Science 575 (2025), https://www.science.org/doi/10.1126/science.adw0018.

[4] *See* e.g., Coenen, C., Grinbaum, A., Grunwald et al. *Quantum Technologies and Society: Towards a Different Spin*. Nanoethics 16, 1–6 (2022). https://doi.org/10.1007/s11569-021-00409-4; Adrian Schmidt et al., *Current and Future Directions for Responsible Quantum Technologies: A ResQT Community Perspective*, arXiv (2025), https://arxiv.org/abs/2509.19815; Pieter E. Vermaas, Luca Possati & Zeki C. Seskir, *Quantum Internet, Governance, Trust, and the Promise of Secure Communication: On Building a Quantum Internet That Will Be Used*, arXiv (2025), https://arxiv.org/abs/2505.15852; and Mauritz Kop & Mateo Aboy et al., *Ten Principles for Responsible Quantum Innovation*, 9 Quantum Sci. & Tech. 035013 (2024), https://doi.org/10.1088/2058-9565/ad3776.

A further implication is that quantum supply assurance and post-quantum cryptography (PQC) migration should be understood as twin pillars of security.[5] The first protects the physical substrate on which mission-relevant quantum capability depends; the second protects long-lived data and critical infrastructure against harvest-now, decrypt-later campaigns. Static national critical-minerals lists are useful but too blunt for this purpose. What is needed is a living, sector-specific criticality method able to register changes in concentration, substitutability, qualification timelines, and geopolitical leverage faster than conventional list cycles permit. Such a method should not remain buried in periodic reports; rather, it should be made visible through a dedicated, continuously updated Quantum Criticality and Critical Minerals (QCCM) dashboard that tracks material concentration, processing exposure, qualification bottlenecks, stockpiling gaps, and geopolitical stress signals across mission-relevant quantum platforms. In policy terms, the dashboard would function as an operational early-warning and prioritization tool: a decision-support interface linking upstream materials risk to downstream deployment consequences in computing, sensing, networking, and secure communications, thereby enabling governments and allied partners to move from static awareness to dynamic resilience management.[6]

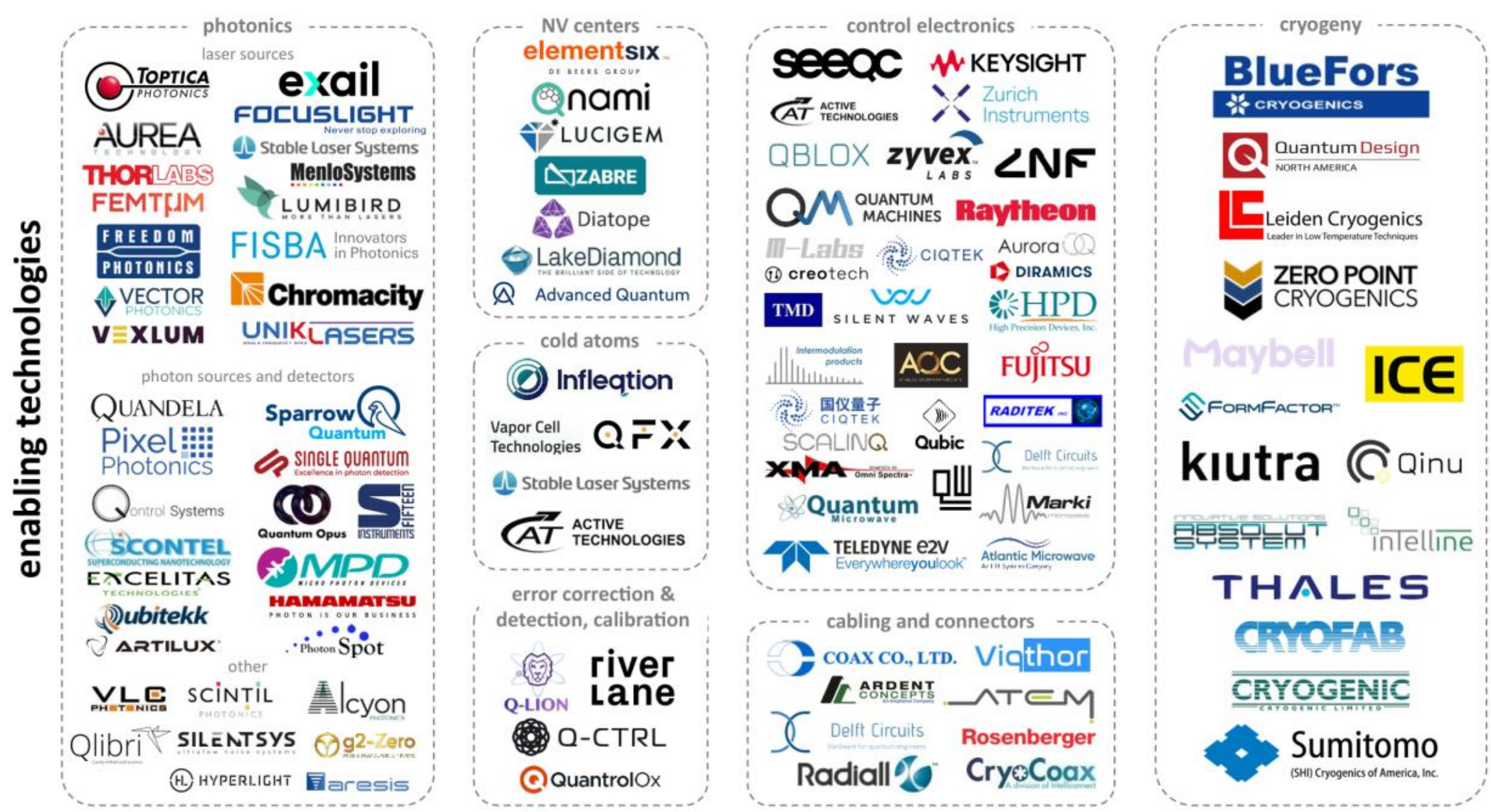

*Figure 1. Differentiating QC from QS for supply-chain criticality (beyond TRL)*[7]

[5] Dongyoun Cho, Mauritz Kop & Min-Ha Lee, *Strategic Governance of Quantum Supply Chains: A Criticality-Based Framework for Risk, Resilience, and Data-Driven Foresight* (Research Square preprint, Nov. 11, 2025), https://doi.org/10.21203/rs.3.rs-8062873/v1.

[6] Mauritz Kop, *A Bletchley Park for the Quantum Age*, War on the Rocks (Nov. 6, 2025), https://warontherocks.com/2025/11/a-bletchley-park-for-the-quantum-age/.

[7] Olivier Ezratty, Understanding Quantum Technologies, p. 646, Sept. 29 (2025), https://www.oezratty.net/wordpress/2025/understanding-quantum-technologies-2025/.

• QC: At scale, QC typically depends on engineered multi-qubit entanglement plus coherent control across many operations (high fidelity, low error rates) in order to realize a true QC advantage. Material and toolchain dependencies are therefore driven by scaling, yield, uniformity, and pathways to error correction.

• QS: Quantum sensing and related quantum platforms often target metrological advantage or thermodynamic advantage where coherence is a primary resource (quantum-state phase accumulation or collective behavior such as superfluidity or superconductivity), while entanglement may be optional, shallow, or localized. Dependencies often shift toward robustness, packaging, calibration, and mission qualification (optics/lasers, vacuum, detectors, and environmental hardening).

• Implication for this study: The same material can matter differently across QC and QS; "criticality" should be tied to the specific entanglement/coherence requirements and the mission profile (including extreme environments), not only to maturity or TRL. For example, the materials requirements of the vast, highly mature laser diode market—perhaps the paradigm case of coherence across quantum technologies—share silicon needs with QC, while the sets of laser and qubit materials are largely disjoint. Figure 1 briefly illustrates the supply chain and related companies for manufacturing various types of quantum computers. It reveals that numerous companies are involved in various fields, including lasers for photon production and control, wafers, electronic control equipment, and cryogenic cooling systems. However, suppliers of the most fundamental and core materials for manufacturing quantum computers or QT devices (such as ultra-high-purity materials, isotopes, shielding materials, and structural materials) are absent. While some materials directly impact quantum technology, such as Rb (rubidium) for neutral atom methods and Tb (terbium) and Er (erbium) for photon methods, remain largely outside the scope of quantum researchers.

## 1. CMM Supply Chains in QT (Use Case 1: Niobium for superconducting QC)

To keep the analysis concrete, we develop two use cases: (i) niobium as a chokepoint material for superconducting QC; and (ii) space SNSPDs as a harsh-environment case that links device degradation to mission and security outcomes. The supply chain for CMMs is fragile. CMMs must be mined and processed into commercially-usable forms or compounds. For a vast number of CMMs, these facilities—mines and processing plants—are owned and operated by a small number of companies located in a limited number of countries. Most of the global supply of indium, for example, is processed in China[8], often recovered as a by-product of zinc refining. Approximately 90% of global niobium mine production comes from Brazil. or rare earths relevant to quantum technologies, such as ytterbium and yttrium, mining (70%) and especially processing (90%) are likewise highly concentrated in China.[9]

[8] U.S. Geological Survey. Indium, Mineral Commodity Summaries 2025, https://pubs.usgs.gov/periodicals/mcs2025/mcs2025 indium.pdf.

Yet the relevant chokepoint is rarely mining alone. Recent mapping studies show that quantum systems depend on a narrow ecosystem of enabling technologies—including dilution refrigeration, cryogenic and vacuum systems, high-purity lasers, photonic components, control electronics, isotopically purified inputs, and single-photon detector materials—many of which are supplied by only a handful of firms or processed outside allied jurisdictions.[10] Alliance-level assessments now suggest that high-purity processing for critical quantum inputs is concentrated overwhelmingly outside NATO territory.[11] Criticality can therefore arise at any layer of the stack: ore, refining, isotopic enrichment, component fabrication, system integration, or maintenance.

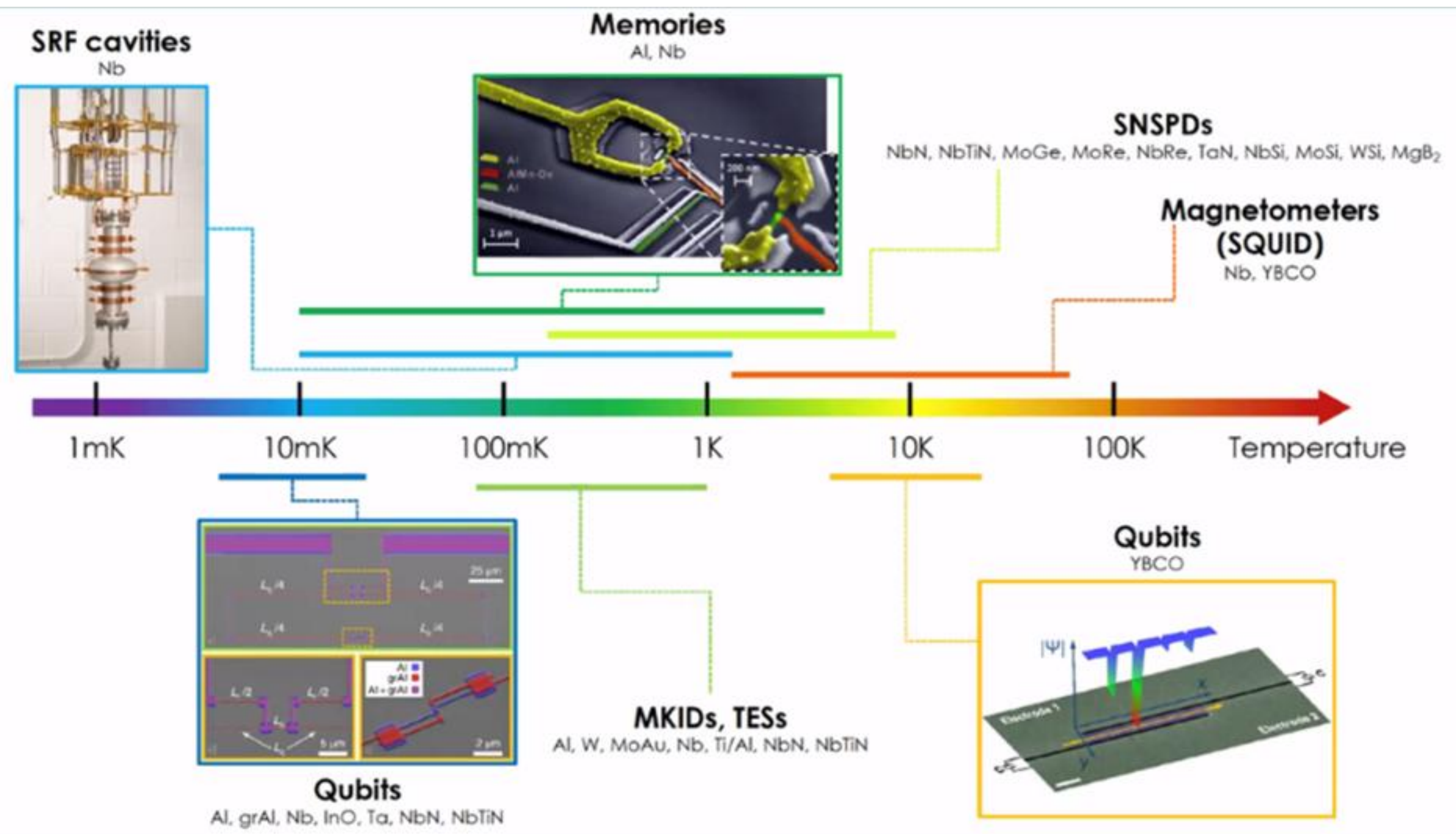

*Figure 2. Superconducting materials (Josephson junction) for QC, Source: DARPA Defense Sciences Office (2023).*

More recently, China has imposed export restrictions on rare earths and other CMMs destined for the United States in response to the Trump administration's tariffs on Chinese imports and

[9] The Washington Post. U.S. rare earth ambitions center on Malaysia. But China's already there, 21 Nov. 2025, https://www.washingtonpost.com/world/2025/11/21/us-china-rare-earth-battle-malaysia/.

[10] Mauritz Kop & Tracey Forrest, *Global Quantum Governance: From Principles to Practice*, CIGI Pol'y Brief No. 222 (Feb. 5, 2026), https://www.cigionline.org/publications/global-quantum-governance-from-principles-to-practice/.

[11] Michal Krelina, *Military and Security Dimensions of Quantum Technologies: A Primer* (Stockholm Int'l Peace Rsch. Inst. 2025), https://www.sipri.org/publications/2025/other-publications/military-and-security-dimensions-quantum-technologies-primer. See also: Lukas Kingma, Freeke Heijman & Carl Williams, *Critical Vulnerabilities in the Quantum Computing Supply Chain within the NATO Alliance* (Official Summary, NATO Transatlantic Quantum Community, May 12, 2025), https://www.fheijman.nl/QSC_report.pdf.

U.S.-led export restrictions to China on advanced node semiconductors and manufacturing equipment. In April 2025, China imposed export licensing restrictions on gadolinium (Gd) and six other heavy rare earth elements (Sm, Tb, Dy, Lu, Sc, Y) for which it holds a virtual monopoly on global supplies.[12] China is also the world's largest producer of gallium (Ga), germanium (Ge), antimony (Sb), and in December 2024, on the heels of new semiconductor-related controls from the U.S. and its allies, China's Ministry of Commerce prohibited their export to the United States.[13] In February 2025, China announced additional controls on bismuth (Bi), indium, molybdenum (Mo), tellurium (Te) and tungsten (W). Even where shipments are not fully blocked, such measures can tighten markets, raise prices, and convert commercial dependence into strategic vulnerability. For example, the price of antimony increased by 250% in the wake of China's controls (for a cumulative price increase of 450% in the last three years) and bismuth rose approximately tenfold at its April 2025 peak, before partially retracing (see Figure 3).[14] The sequencing is analytically important: the February 2025 licensing regime, not the April 2025 reciprocal-tariff announcement, is the proximate driver of the bismuth dislocation shown in Figure 3. This is a case study in how a monopsonist supplier can convert commercial dependence into pricing leverage through administrative, rather than tariff, instruments.

[12] Reuters. China hits back at US tariffs with export controls on key rare earths, 4 April 2025, https://www.reuters.com/world/china-hits-back-us-tariffs-with-rare-earth-export-controls-2025-04-04/.
[13] Reuters. China's Export Ban Push Antimony Prices to New Highs, 6 Jan. 2025, https://www.reuters.com/markets/commodities/chinas-export-ban-push-antimony-prices-new-highs-2025-01-06/.
[14] Min-Ha Lee, *Critical Materials for Next Generation: Quantum Information Technologies*, (Critical Materials Innovation Hub US DOE annual meeting, Idaho National Laboratory, Aug. 28, 2025).

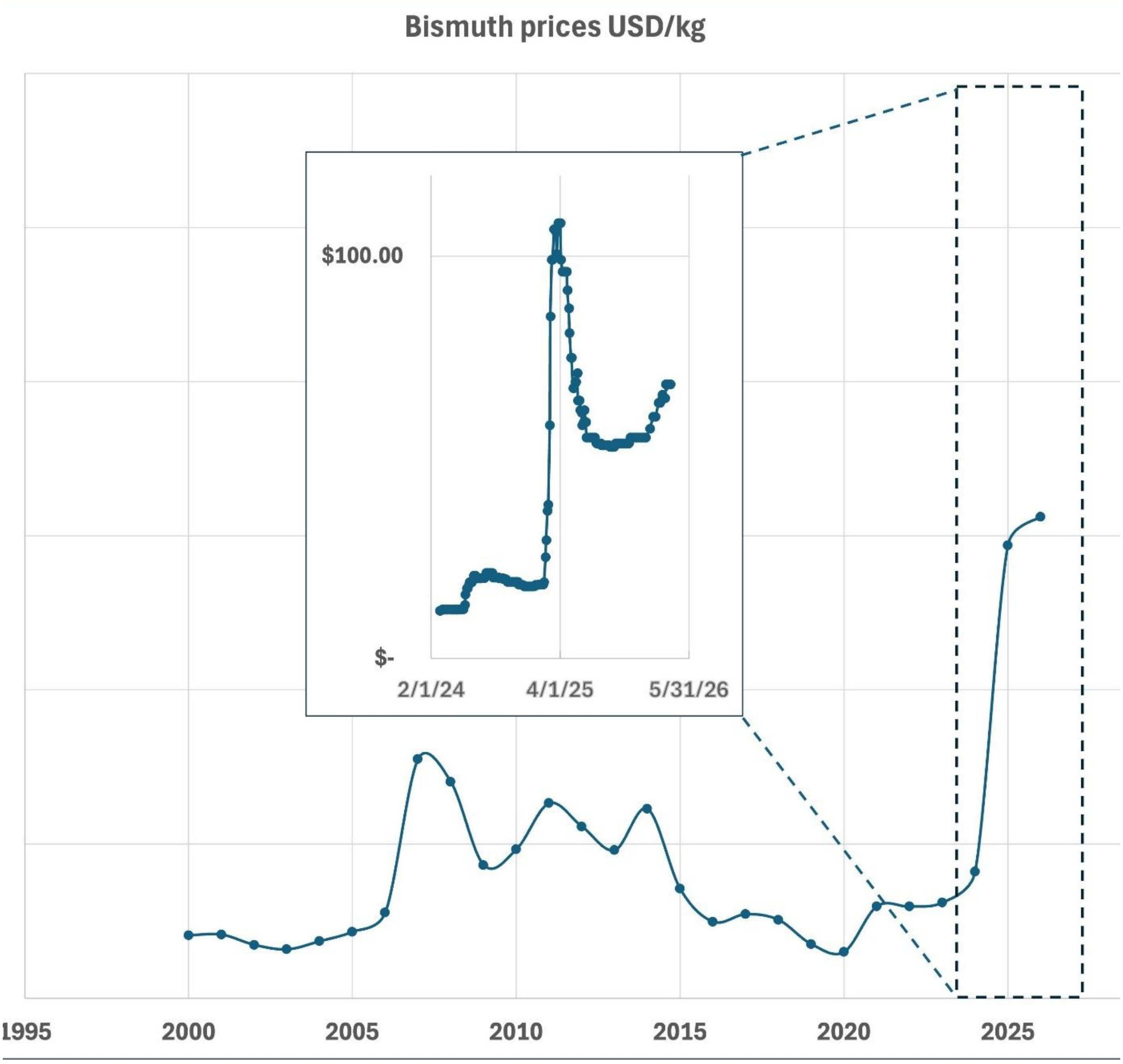

*Figure 3. Bismuth price, 2006–2026 ($/kg). USGS annual dealer prices (2006–2020, converted from $/lb at 2.2046 lb/kg) and Strategic Metals Invest spot observations (2021–2026) show a step change following China's 4 February 2025 addition of bismuth to its dual-use export control list, with the spot price peaking at $108.15/kg on 2 April 2025 — roughly an order of magnitude above the 2021–2024 baseline of approximately $12/kg. The 2020–2021 discontinuity reflects a data-source change (USGS dealer price to SMI spot), not a price event; USGS dealer quotations run systematically below private-sector spot by roughly 30–60% under normal conditions.*

A clear precedent emerged in 2010, when China restricted rare earth exports to Japan during the Senkaku Islands dispute. That episode demonstrated that concentrated mineral supply chains can be weaponized for geopolitical purposes. Since then, concern over critical-minerals vulnerability has spread across the United States, Europe, Japan, and other import-dependent economies, and recent tariff and export-control escalation has renewed those concerns.[15]

[15] T. Erdmann, Diversifying the U.S. Rare Earth Element Resource Base is Vital to National Security, U.S. Naval War College Report, 31 Mar. 2023, https://apps.dtic.mil/sti/html/trecms/AD1209298/index.html

In this respect, quantum supply chains are best understood not as a narrow mining problem but as a test case for a more fragmented geoeconomic order in which states increasingly treat refining, processing, and downstream industrial ecosystems as instruments of leverage. Recent commentary on China's rare-earth posture suggests that Beijing is now more willing than in earlier trade disputes to use its accumulated dominance as recurring leverage,[16] while broader strategic analysis places quantum supply chains squarely within the reordering of global economic interdependence.[17]

The structural problem is broader than any single material. China is the leading producing country for a large share of globally critical minerals, while the United States remains heavily import-reliant across much of the critical-minerals portfolio. USGS data show that the United States is 100% net import-reliant for a subset of listed critical minerals (13 CMMs) and more than 50% import-reliant for many others (31 CMMs).[18] Production is also highly concentrated: for several critical minerals, one country accounts for more than half of global output, and in some cases far more. Niobium is especially salient in this respect because supply is geographically concentrated, substitutes are limited for certain high-performance applications, and demand is rising across advanced industrial and quantum uses.

Accordingly, the relevant question is not only whether a material is technically important, but whether its supply can be diversified, substituted, stockpiled, or otherwise mitigated at the speed required by mission needs. In the case of superconducting QC, niobium illustrates how a material can become strategically critical when performance requirements are stringent, substitutes are imperfect, and supply remains concentrated in a narrow geopolitical and industrial base.

## 2. Stockpiling

The European Union, Japan, South Korea, the United States, and other importers have sought to reduce their exposure to these risks by stockpiling reserves of certain CMMs, pursuing multilateral partnerships, and incentivizing the development of a diversified base of suppliers (including domestic mining and processing) with subsidies and other market interventions. Broadly speaking, these efforts are risk-informed. For example, in 2025, the U.S. Geological Survey (USGS) published[19] a statutorily mandated list of fifty CMMs to guide U.S. efforts to improve resilience against supply disruptions, and the U.S. Department of Defense's National Defense Stockpile (NDS) prioritizes stockpiling of CMMs [20] to sustain military readiness.

[16] Kop & Forrest, *supra* note 10.
[17] Mauritz Kop, *The Nexus of Quantum Technology, Intellectual Property, and National Security: An LSI Test for Securing the Quantum Industrial Commons*, arXiv:2602.15051 (2026), https://arxiv.org/abs/2602.15051.
[18] U.S. Geological Survey. Mineral Commodity Summaries 2026, https://pubs.usgs.gov/periodicals/mcs2026/mcs2026.pdf.
[19] U.S. Geological Survey. Mineral Commodity Summaries 2025, https://pubs.usgs.gov/periodicals/mcs2025/mcs2025.pdf.
[20] Defense Logistics Agency. Strategic Materials: About, https://www.dla.mil/Strategic-Materials/About/. Accessed 9 June 2025.

Stockpiling, however, is only as good as the prioritization logic behind it. Broad national lists provide an indispensable macroeconomic baseline, but they can miss quantum-specific dependencies that are small in tonnage yet decisive in system effect—helium-3, silicon-28, rubidium isotopes, detector materials, and other niche inputs with years-long qualification timelines. Recent criticality work argues that quantum policy therefore needs a higher-resolution instrument that separates raw-material concentration from refining concentration, commercial inconvenience from mission failure, and near-term substitutability from technically possible but operationally unrealistic redesign.[21]

While stockpiling is considered sound policy to mitigate nation-level supply risk, enduring action has been difficult to maintain in the United States.[22] The 2026 U.S. initiative, Project Vault, for stockpiling critical minerals and materials, which mentions fifty materials, reprises historic efforts. Informal stockpiling began during World War I in 1917; one of the first national stockpiles was the Federal Helium Reserve, created in 1925 in anticipation of a need for military dirigibles. The first materials program was created in 1939 and revised often through World War II and the Cold War.[23] Inevitably, budget pressures arose prompting the US Congress to start selling off federal stockpiles in the 1960s to pay for Cold War expansion. The sell-off continues to this day: for example, the Federal Helium Reserve, repurposed from dirigibles to flushing liquid-fueled rocket motors, was finally sold off completely in 2024.[24]

By statute, fuel materials, gases, and manufactured elements are not included in the set of critical minerals and materials. Hence, neither the Strategic Petroleum Reserve nor the Federal Helium Reserve are addressed in Project Vault or in the USGS critical minerals and materials lists.[25] Man-made isotopes including medical ones are uniformly and largely ignored in almost all discussions, however uranium has been identified as critical in the 2025 USGS list after significant public comment, but not as a fuel element.

Notably, QC, QS, and certain energy technologies such as fusion depend on several isotopically pure elements. Examples include natural helium-4 and its man-made isotope helium-3; closely related nuclei are hydrogen-2 (deuterium), hydrogen-3 (tritium), and lithium-6 and -7. These light elements are famously used in national security applications and certain medical settings; less appreciated are their roles in QC/QT, especially dilution refrigeration. Lithium-6 and -7 are important as targets for producing tritium, which decays to helium-3; there is no other practical pathway to helium-3.

---

[21] *See* Min-Ha Lee, *A Framework for Assessing Vulnerabilities in the Quantum Computing Materials Supply Chain* (GTG Policy White Paper, Stanford Univ., Oct. 15, 2023), https://cisac.fsi.stanford.edu/publication/framework-assessing-vulnerabilities-quantum-computing-materials-supply-chain; and Farrell Gregory, *China's Rare Earths Chokehold: A Primer*, ChinaTalk (Jan. 8, 2026), https://www.chinatalk.media/p/chinas-rare-earths-chokehold-a-primer.

[22] Nat'l Rsch. Council, Managing Materials for a Twenty-First Century Military (Nat'l Acads. Press 2008), https://doi.org/10.17226/12028.

[23] Id. at 133.

[24] *See* e.g., BLM Press, BLM completes sale of Federal Helium System, Jun 27, 2024, https://www.blm.gov/press-release/blm-completes-sale-federal-helium-system.

[25] Export-Import Bank of U.S. Project Vault and the U.S. Strategic Critical Mineral Reserve, Feb. 6 (2026), https://www.exim.gov/news/week-review-project-vault-and-strategic-critical-mineral-reserve.

For superconducting QC, this dependency is operationally significant: a large cryostat can require on the order of 5 moles of a helium-3/helium-4 mixture to sustain ultralow millikelvin temperatures at high power load, whereas current helium-free magnetic cooling remains limited to roughly 100 mK and therefore does not eliminate dependence on isotopic helium for the coldest regimes.

Heavier nuclei are also important for QT research, for example silicon-28 as a substrate. Quantum metrology and some qubit architectures likewise depend on rubidium isotopes, including Rb-87. Medical, quantum, and materials researchers routinely require virtually every stable isotope and even some radioactive ones. For example, technetium-99 (derived from its parent molybdenum-99) is used in millions of medical procedures every year.[26]

A potentially rich source of heavy nuclei consists of spallation targets stored at large federal accelerators such as LANSCE (Los Alamos Neutron Science Center) at Los Alamos and SNS (Spallation Neutron Source) at Oak Ridge. Often buried on-site as high-level radioactive waste, these targets—typically tungsten or mercury—contain substantial inventories of exotic nuclei, potentially "mine-able" if someone wants to pay for the chemical and physical separations.

The Defense Logistics Agency Strategic Materials (DLA Strategic Materials) is responsible for the operational oversight of the National Defense Stockpile (NDS) of strategic and critical materials. Managing stockpile security, providing environmentally sound stewardship, and ensuring the readiness of all NDS stocks are central parts of the mission of DLA Strategic Materials. The NDS currently contains fifty-two unique commodities stored at nine locations within the continental United States.[27]

Recent U.S. actions suggest a shift from passive monitoring to active pre-positioning. The January 2026 White House proclamation on processed critical minerals treats processed inputs and their derivatives as national-security concerns and emphasizes negotiated supply assurance,[28] while Project Vault reflects a more explicit turn toward reserve formation, public-private financing, and industrial pre-positioning for strategic manufacturing.[29] In the quantum context, that matters because the most consequential vulnerability often lies not in raw ore but in processed, qualified, and deployable forms of material.[30]

The NDS is the defense sector's answer to the supply-risk problem: it prioritizes a subset of CMMs with critical defense applications and calibrates the size of the NDS stockpile for a given material to defense needs, buying and selling materials primarily to support warfighting.[31] Since other strategically important sectors likely face a different constellation of dependencies on CMMs–for example the medical community's reliance on superconducting magnets for Magnetic Resonance Imaging (MRI)–a successful, ethical strategy to hedge against supply disruptions must consider sector-specific factors as a starting point for specifying aggregate resiliency requirements.

[26] *See* e.g., Antonino P., Marco C. & Lina Q., Nature 603, 393 (2022), Secure molybdenum isotope supplies for diagnostics, https://www.nature.com/articles/d41586-022-00705-3.
[27] Defense Logistics Agency, *supra* note 20.
[28] The White House, Adjusting Imports of Processed Critical Minerals and Their Derivative Products into the United States (Proclamation) (Jan. 14, 2026), https://www.whitehouse.gov/presidential-actions/2026/01/adjusting-imports-of-processed-critical-minerals-and-their-derivative-products-into-the-united-states/.
[29] Alan Rappeport & Tony Romm, *Trump Unveils $12 Billion Critical Minerals Stockpile*, N.Y. Times (Feb. 2, 2026), https://www.nytimes.com/2026/02/02/business/trump-critical-minerals-stockpile.html. Export-Import Bank of U.S. Project Vault and the U.S. Strategic Critical Mineral Reserve, Feb. 6 (2026), https://www.exim.gov/news/week-review-project-vault-and-strategic-critical-mineral-reserve.
[30] Min-Ha Lee, *Quantum Criticality Index: Understanding Critical Raw Materials Supply Chains in Quantum Technologies* (Stanford Ctr. for Responsible Quantum Tech. project, 2024).

The Department of Defense maps the relationships and interdependencies across the entire U.S. economy for CMMs over multiyear periods, and a separate U.S. interagency effort in partnership with U.S. national laboratories has conducted an apparently similar mapping focused more narrowly on rare earth elements. Both efforts are resource intensive and approach supply chain vulnerability from a macroeconomic perspective in an attempt to understand systemic risk.

The stakes are high: whoever gets there first will be in a commanding position to corner the market for modern technologies and serve as the seed bed for new industries and scientific discoveries made possible by the staggering computational potential of quantum mechanics.

Whoever gets there first will also have powerful economic and national security reasons to limit which countries get access to the full powers of QT. In addition to reaping the potentially enormous economic benefits of hosting an upstart "Silicon Valley" for QC, the winner could also potentially break the encryption algorithms used globally to protect the integrity of trillions of dollars in transactions and to safeguard personal, business, and governmental secrets. A prediction by the majority of quantum executives surveyed in 2022 [32] that the sector would suffer a supply chain disruption by 2025 has proven prescient, as China's coercive use of export controls on CMMs demonstrates. A tit-for-tat dynamic between China and the U.S. on tariffs and export controls has already impacted the critical raw materials supply chain, with more disruptions possible in the future—including controls targeting the QC supply chain.[33]

---

[31] C. Keys, et al., Emergency Access to Strategic and Critical Materials: The National Defense Stockpile, Congressional Research Service report R47833 (Nov. 14, 2025), https://www.congress.gov/crs-product/R47833.

[32] B. Sorensen and T. Sorensen, Challenges and Opportunities for Securing a Robust US Quantum Computing Supply Chain, QED-C, June 2022, https://quantumconsortium.org/publication/quantum-computing-supply-chain-issues/.

[33] *See* also Peter Alexander Earls Davis, Mateo Aboy & Timo Minssen, *Regulatory Challenges and Opportunities of Export Controls on Quantum Computing*, in *Quantum Technology Governance: Law, Policy & Ethics in the Quantum Era* (Mateo Aboy, Marcelo Corrales Compagnucci & Timo Minssen eds., forthcoming), available at SSRN (Aug. 2025), https://ssrn.com/abstract=5404548.

| Facility | Date | Operation company | Material | Location | Feature | Note |
|---|---|---|---|---|---|---|
| Ge production | 2024-4 | DoD | Ge | **Utah**/ Saint George | High-purity germanium wafers of semiconductor grade 5N or higher | The U.S. Department of Defense is investing $14.4 million in a Utah manufacturing facility under the Defense Production Act (DPA) to produce high-purity germanium for satellites and solar panels over the next four years. |
| Cu, Te refining, smelting | 2025-7 | Kennecott Utah Copper LLC; a division of Rio Tinto Group | Te | **Utah**/ Slat Lake city, Bingham Cayon Copper mine | Extraction of tellurium from mine waste | Freeport-McMoRan, a US mining company, plans to extract copper from mine waste, coal waste, and abandoned uranium mines under the Defense Production Act (DPA). |
| U, HREE refining, smelting | 2025-7 | Energy Fuels Co. | Dy, Tb, Sm (HREE) | **Utah**/ White Mesa Mill facility | Extraction of heavy rare earth oxides from rare earth ores | Currently Dy, Tb completed, Sm in progress, pilot scale commercial production planned for the second half of 2026 |
| Ni, Li, REE refining, smelting | 2025-6 | Westwin Elements Co. | Ni, Li, REE | **Oklahoma**/ Lawton | Nickel refining, Recycling battery | It has an annual production capacity of 200 metric tons and is expected to expand to 34,000 metric tons by 2030, which is enough to meet approximately 10% of the total nickel demand in the United States. |
| Zn, Ge refining, smelting | 2025-7 | Freeport-McMoRan Co. | Ge | **Oklahoma**/ Tar Creek Mine (abandoned) | Extraction of germanium from mine waste | Freeport-McMoRan, a US mining company, will prioritize securing highly import-dependent minerals such as zinc, germanium, and rare earth elements from mine waste, coal waste, and abandoned uranium mines - based on the Defense Production Act (DPA). |
| Zn, Ga, Ge refining | 2025-9 | Nyrstar; a division of Trafigura Group | Ga, Ge | **Tennessee**/ Clarksville mine | Gallium and germanium processing facilities under construction, nickel and lithium facilities planned | Recently, the company invested $150 million in expanding its facilities for the recovery and processing of rare metals, including gallium and germanium, through advanced technologies. This will support annual production of 40 tons of gallium and 30 tons of germanium at its zinc smelter. The company plans to produce semiconductor-grade gallium and high-purity germanium within the United States, meeting 80% of domestic demand. |
| REE oxides separation | 2025-9 | Ucore Rare Metals Inc | REE | **Louisiana**/Alexandri a | Establishment of a rare earth separation facility | US$22.4 million funding agreement with the US Army Contracting Command-Orlando to launch its RapidSX™ rare earth element separation technology; RapidSX™ full commercial scale-up engineering and testing program |
| Cu, Ga, REE refining | 2025-9 | Alliance Resource Partners L.P. | Ga, REE | Illinois and 7 states, **Australia** | Copper, gallium and rare earths production | Investing over $600 million in overseas mine development in partnership with the U.S. government's International Development Finance Corporation (DFC). As the second-largest coal producer in the eastern U.S., it operates seven underground mine complexes in the Illinois Basin and Appalachia. |
| REE refining, smelting | 2025-9 | US Strategic Metals Co. | REE | **Pakistan** | Establishment of a rare earth refinery | Decided to invest $500 million in core mineral development projects |
| Sb refining, smelting | 2025-12 | Perpetua Resources Co. | Sb | **Idaho** | Establishment of small scale portable refineries | Perpetua Resources Partners with the Idaho National Laboratory via Battelle Energy Alliance LLC to Advance Critical Mineral Pilot Plant for intending to demonstrate the feasibility of producing high-quality, military specification antimony trisulfide |

*Table 1. U.S. government projects to resolve criticality of CMMs.*[34]

Table 1 contains domestic and international projects related to mitigatability of CMMs currently being led by the U.S. government. Mining, smelting, and refining projects for rare earth elements, gallium, nickel, and antimony are actively underway. One point to note is that while these movements are not new and have long been considered important, action plans are failing to keep up with global trends regarding these key materials. For example, regarding germanium, in 2012 the Department of Defense had been pursuing a project to produce high-purity germanium wafers of semiconductor grade 5N or higher (these high purity materials are also essential in quantum technologies as base materials) at its own production facility.[35] However, due to sluggish progress, in 2024, the U.S. DoD is to invest an additional $14.4 million in a Utah manufacturing facility under the Defense Production Act (DPA) to produce high-purity germanium for satellites and solar panels over the next four years.[36]

[34] Min-Ha Lee, *Addressing vulnerability of rare material supply chains to support national security and critical materials manufacturing*, (Center for Advanced Vehicular Systems (CAVS) Seminar, Mississippi State University, (Dec. 16, 2025).

[35] 5N Plus Subsidiary Awarded U.S. Defense Logistics Agency Strategic Materials Contract, CISION (Oct. 19, 2012), https://www.newswire.ca/news-releases/5n-plus-subsidiary-awarded-us-defense-logistics-agency-strategic-materials-contract-510886821.html.

[36] DOD Awards $14.4 Million to Sustain and Enhance the Space-Qualified Solar Cell Supply Chain (Apr. 13, 2024), https://www.war.gov/News/Releases/Release/Article/3743467/dod-awards-144-million-to-sustain-and-enhance-the-space-qualified-solar-cell-su/.

| Indicator | Operational meaning / proxy | Why it matters for criticality |
|---|---|---|
| Supply concentration (mining) | High share of extraction in 1–2 jurisdictions; limited alternative deposits | Elevates disruption risk and limits rapid diversification |
| Supply concentration (processing/refining) | Processing dominated by a small number of firms/jurisdictions; limited spare capacity | Creates chokepoints even when raw ore is available |
| Trade and export controls / geopolitical exposure | Active or plausible restrictions, sanctions, or licensing regimes affecting supply | Transforms commercial dependence into strategic vulnerability |
| By-product dependence | Material primarily recovered as a by-product of another metal; recovery tied to host-metal economics | Constrains ramp-up and complicates resilience planning |
| Low substitutability for required performance | Few substitutes without major performance loss (e.g., superconducting properties; optical/detector properties) | Limits mitigations via redesign or material substitution |
| Long lead times and qualification barriers | Years-scale timelines for permitting, processing expansion, or mission qualification | Makes short-notice response impractical |
| Limited recycling/stockpiling practicality | Low recycling rates; small, specialized markets; storage/handling constraints | Reduces buffer options against shocks |
| Single points of failure in toolchain or BOM | Dependency on specific sputtering targets, wafers, cryo-components, lasers, or detectors | Creates systemic fragility beyond the material itself |

*Table 2. Vulnerability indicators for quantum-relevant CMMs: operational proxies and criticality rationale (Criticality Method).*

## 3. Use Case 1: Niobium for Superconducting QC and Nickel for Mu-Metal Shielding

### 3.1 Niobium

Niobium (Nb) is a metal characterized by its high melting point as a refractory metal, high electrical conductivity, ductility, high strength, and exceptional resistance to corrosion.[37] These extraordinary properties have rendered niobium indispensable across a broad spectrum of industrial and technological applications.[38]

Serving as a microalloying element in steel, for instance, the Øresund Bridge in Sweden utilized steel containing 0.02 wt.% Nb, leading to a 15% weight reduction and cost savings of 25 million USD.[39] Furthermore, niobium plays a crucial role in the production of superalloys and is of significant importance in aerospace and power-generation technologies.[40] Its exceptional conductive properties also find applications in the healthcare industry, such as in MRI machines.[41] Niobium is increasingly relevant to post-Moore computing architectures and is already a key input for superconducting quantum platforms.

[37] Dolganova, I., Bosch, F., Bach, V., Baitz, M., Finkbeiner, M., Life cycle assessment of ferro niobium. Int. J. LCA 25 (3), 611–619 (2020).
[38] Moisés, G, Jinhui L., Xianlai Z., Niobium: The unseen element - A comprehensive examination of its evolution, global dynamics, and outlook, Res. Cons. Recycl. 202, 107744 (2024).
[39] Bakry, M., Li, J., Zeng, X., Evaluation of global niobium flow modeling and its market forecasting. Frontiers in Energy 17 (2), 286–293 (2023).

We define these CMMs as "Critical Level I" materials. The rule-based decision-flow model shows schematically how we can evaluate the degree of vulnerability for quantum-relevant materials, components, and equipment using the vulnerability.[42] Table 3 provides an illustrative, non-exhaustive list of candidate "Critical Level I" materials for which supply and/or processing is highly concentrated and mitigatability (substitution/design-around/stockpiling) is limited for specific quantum applications. Recent Chinese measures illustrate how trade restrictions can rapidly translate technical dependence into strategic vulnerability: on Dec. 3, 2024, China announced a ban on exports of gallium, germanium, and antimony (and certain superhard materials) to the United States; on Feb. 4, 2025, China announced export controls (licensing) on items related to tungsten, tellurium, bismuth, indium, and molybdenum. These are used here as a stress test of the trade-restriction/geopolitical-exposure indicator rather than as a claim that any given material is irreplaceable across all quantum platforms.

This layered picture has direct governance implications. Recent work on quantum export controls shows that the object of regulation is not a single "quantum computer," but a modular stack spanning materials, cleanroom tools, cryogenics, software, cloud access, and specialized know-how.[43] For materials such as niobium and nickel-based shielding inputs, the policy challenge is therefore capability-targeted and stack-aware: the goal is not indiscriminate denial, but calibrated measures that protect genuinely sensitive chokepoints without freezing benign research collaboration, allied scale-up, or standards development.

For example, transition metals titanium and niobium are used in superconducting cables; niobium is also used in superconducting Josephson qubits. The transition metals iron, cobalt, nickel, and chromium are used in cryostats. Helium is used in cryostats operating below 10 degrees Kelvin, and helium-3 is used to reach temperatures below 3 degrees Kelvin. Nitrogen is also used in some cryostats, mostly for QS. Carbon is used in nanotubes for silicon qubits, which in turn use an isotope of silicon (silicon-28) for the silicon-qubit wafers. Silicon is also used in wafers for spin, electron, and photonic qubits. Copper, silver, and gold are used in cryostats for cold-plate cabling. The rare earths ytterbium, europium, praseodymium, erbium, and neodymium are used in trapped-ion qubits, optical memories, and lasers.

---

[40] Heisterkamp, F., Carneiro, T., Niobium: Future possibilities - Technology and the Market Place. Niobium, Science and Technology (2011).
[41] McCaffrey, D.M., Nassar, N.T., Jowitt, S.M., Padilla, A.J., Bird, L.R., Embedded critical material flow: The case of niobium, the United States, and China. Resources, Conservation and Recycling 188, 106698 (2023)
[42] Cho, Kop & Lee, *supra* note 5.
[43] Ulrich Mans, *Quantum Supply Chains: A Test Case for a New Economic World Order*, Just Security (Aug. 27, 2025), https://www.justsecurity.org/115813/quantum-new-economic-world-order/.

Superconducting qubits are electronic circuits built with techniques akin to those used to manufacture legacy analog circuits for the radar and electronics markets, with further similarities to the CMOS processes used to manufacture semiconductors. Materials used for manufacturing superconducting qubits include aluminum, boron (in the form of boron nitride), and titanium (in the form of titanium nitride) for the dielectrics used in the Josephson junction; niobium for capacitors, resonators, and sometimes the Josephson junction; indium for the chipset connectors; selenium and boron for capacitors; silicon or sapphire for the wafer substrate; and tantalum for superconductors and resonators. Quantum-computing qubit research, development, and commercialization depend on having access to these materials or a viable substitute, if one exists.

Niobium (Nb) serves as an alloying agent in high-strength low-alloy steel, which is ideal for bridges, skyscrapers, oil pipelines, vehicles, superconductors, and quantum technologies as well. Niobium is not mined in either China or the United States. The U.S. net import reliance of niobium in 2025 is 100% (67% from Brazil, 28% from Canada, 5% from others).[44] Brazil has dominated the production of niobium, accounting for nearly 90% of global production since the 1990s, with approximately 85% of world primary Nb production coming from a single mine in Brazil, with Canada following as the second-largest producer, contributing approximately 8.6%. [45]

Nevertheless, due to niobium's heavy reliance on two countries for mine production (97.3% in 2024), its growing integration into modern, energy-efficient, low-emission technologies, and the absence of viable substitutes without substantial performance trade-offs, niobium has been classified as a critical material by the European Union,[46] the United States,[47] Australia,[48] India,[49] Japan, and Canada.[50]

The dominant player in this sector is Companhia Brasileira de Metalurgia e Mineração (CBMM), responsible for 81% of the world's niobium production from domestic mines. In a strategic move, China Nonferrous Metal Mining Group (CNMC) acquired Mineração Taboca S.A. The deal, finalized on November 26, 2024, grants China access to vital raw materials for green and electronic technologies. Taboca, Brazil's largest refined tin producer, has output that extends beyond tin. It also produces niobium, tantalum, and rare earth elements, crucial components in electronics manufacturing. FeNbTa (iron-niobium-tantalum) finds wide application in the chemical industry. It is used to create products for the electronics, aerospace, and medical-device industries. CNMC, already the world's largest copper producer with operations in Zambia, is diversifying its portfolio with this purchase. This acquisition aligns with China's strategy to secure raw materials for its growing technology sector. It also reflects the increasing global demand for rare earth elements and other strategic minerals. [51]

[44] U.S. Geological Survey. Mineral Commodity Summaries 2026, *supra* note 18.
[45] A. L. Gulley et al., China, the United States, and competition for resources that enable emerging technologies, PNAS, 16, 4111 (2018); www.pnas.org/cgi/doi/10.1073/pnas.1717152115.
[46] Grohol, M., Veeh, C., GROW, D., European Commission, Study on the Critical Raw Materials for the EU 2023 – Final Report. Belgium, p. 160 (2023), https://single-market-economy.ec.europa.eu/publications/study-critical-raw-materials-eu-2023-final-report_en.
[47] United States Geological Survey (USGS), Niobium and Tantalum Statistics and Information (2026), https://www.usgs.gov/centers/national-minerals-information-cente r/niobium-and-tantalum-statistics-and-information.
[48] Australian Government, Department of Industry, Science and Resources. Critical Minerals Strategy, 2023–2030 (2023); https://www.industry.gov.au/publications/critical-minerals-strategy-2023-2030.
[49] India. Ministry of Mines, Critical minerals for India (2023), https://mines.gov.in/admin/download/649d4212cceb01688027666.pdf.
[50] Graedel, T.E., Reck, B.K., Miatto, A., Alloy information helps prioritize material criticality lists. Nat. Commun. 13 (1), 150(2022), https://www.nature.com/articles/s41467-021-27829-w

A related development occurred earlier. In 2016, a US$1.5 billion deal for Brazilian niobium and phosphate mines marked the expansion of China's China Molybdenum Company Limited (CMOC) and the contraction of Anglo American. In one of the most important transactions in the Brazilian mining sector, CMOC acquired control of large Brazilian niobium and phosphate mines. CMOC beat off competition from at least fifteen other companies to acquire 100% of Anglo American's niobium and phosphates operations in Brazil. The acquisition positioned CMOC as the world's second-largest producer of niobium and one of the main suppliers of phosphate fertilizers in the world. The asset package acquired by CMOC was robust and included the Boa Vista Mine, in Catalão (GO), one of the largest niobium operations in the world. The purchase of the niobium business was seen as an important strategic addition because the metal is a critical input for the production of special steels, vital to China's high-tech industry. After purchasing the Brazilian niobium mine, CMOC increased niobium production by more than 50%, reaching historical records. Those operations, located in Catalao and Ouvidor in the state of Goias and supported by processing facilities in Cubatao in Sao Paulo state, made CMOC the second-largest producer of niobium globally.[52]

In short, Chinese firms have pursued a coordinated strategy of overseas mineral acquisitions, including niobium assets in Brazil. As an example, in 2011 and 2016 Chinese state-owned enterprises (SOEs) acquired a 15% equity share of Brazil's largest mine and 100% of the second Brazilian mine. While these investments will not increase global niobium supplies, they will likely reduce the risk of supply disruptions for China while simultaneously increasing China's strategic leverage over niobium and increasing the risk of supply disruptions to countries outside China.[53]

Because niobium sits at the intersection of metallurgy, superconducting performance, and geographically concentrated mining, it is a useful example of a material that may be economically modest in headline market size yet strategically outsized in downstream effect.[54]

---

[51] The Rio Times, Chinese Mining Giant Secures Brazilian Rare Earth Company, November 28, (2024), https://www.riotimesonline.com/chinese-mining-giant-secures-brazilian-rare-earth-mining-company/.

[52] Barbara Lewis, *Anglo American to Sell $1.5 Bln of Brazilian Assets to China*, Reuters (Apr. 28, 2016), https://www.reuters.com/article/business/anglo-american-agrees-15-billion-sale-of-brazilian-niobium-phosphates-units-idUSKCN0XP15S/.

[53] Click Oil and Gas, June 25 (2025), https://en.clickpetroleoegas.com.br/In-2016--a-US%2417-billion-deal-for-Brazilian-niobium-and-phosphate-mines-sealed-the-dispute-between-the-expansion-of-the-Chinese-company-CMOC-and-the-contraction-of-the-Anglo-American-company-MHBB01/.

[54] Lee, *supra* note 21.

That mismatch between tonnage and strategic importance is precisely why general minerals lists and volume-based industrial-policy instruments can miss quantum-relevant chokepoints.[55]

Considering the relevance and localized production of niobium in the world, even though it remains largely unseen, it is rather surprising that its strategic importance has not attracted greater policy attention in user countries. Given its significance for numerous governments, this trend comes as China expands its grip on South America's resource and CMM sector even as the U.S. administration earmarks the Western Hemisphere as its principal sphere of influence.

While the notion of niobium criticality arises as a potential concern, possible substitutes include refractories such as its close counterpart tantalum (Ta), or elements like molybdenum (Mo), titanium (Ti), tungsten (W), and vanadium (V) depending on the application, although substitution may entail significant performance trade-offs.

### 3.2 Nickel

For magnetic shielding of superconducting quantum computers, it is important to use a metal with high relative magnetic permeability (greater than or equal to 100,000). These materials are often referred to as MuMetal®, μ-metal, or they may have specific names, such as Cryoperm, Cryophy, or Amumetal[56] Shields made of ferromagnetic materials, such as permalloys (μ-metal, a soft magnetic alloy, or steels), with high relative magnetic permeability are used to protect sensitive equipment like superconducting quantum computers, MRI systems, and transformers from static and slowly varying (up to 1kHz) magnetic fields. μ-metal is a particular permalloy, an iron-nickel alloy with 77~80% nickel content. The high nickel content leads to a relative permeability above 100,000 and a coercive field of about 1 A/m. This permeability value allows better shielding performance in comparison with other ferromagnetic materials. A small number of additional elements such as molybdenum, silicon, and manganese is also added to stabilize the material lattice. These elements constitute a few percent of the material composition.[57]

---

[55] Gregory, *supra* note 21.

[56] E. I. Malevannaya et al., An engineering guide to superconducting quantum circuit shielding, Appl. Phys. Rev. 12, 031334 (2025), https://doi.org/10.1063/5.0250262.

[57] P. Arpaia et al., Magnetic characterization of Mumetal for passive shielding of stay fields down to the nano-Tesla level, Nuclear Inst. and Methods in Phys. Res. A988, 164904 (2021), https://doi.org/10.1016/j.nima.2020.164904.

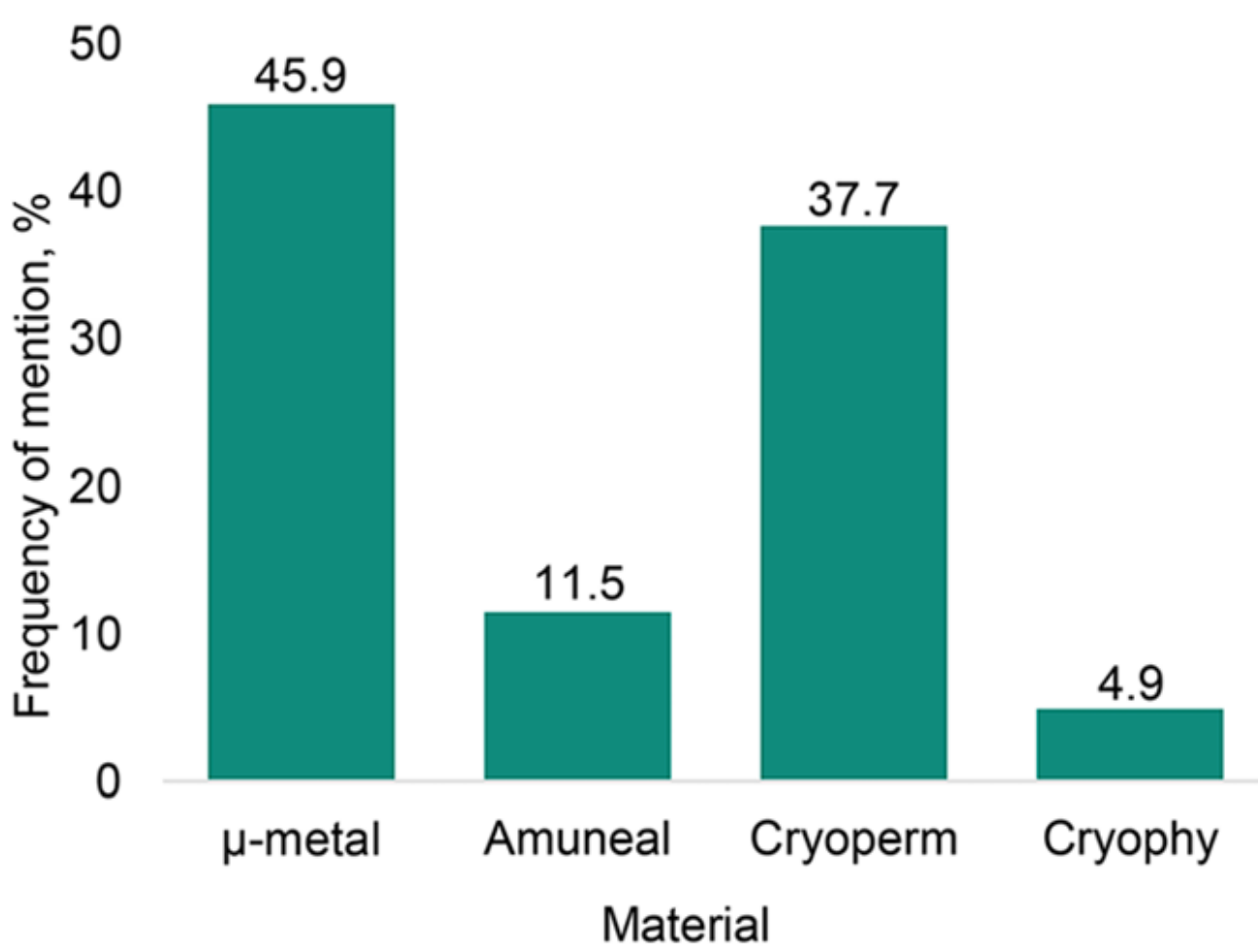

*Figure 4. Typical materials used for magnetic shields.*[58]

Niobium is central to superconducting qubit architectures, while nickel is central to the shielding environment those architectures require. Nickel (Ni) serves as a crucial alloying element in stainless steel, superalloys, and powders, which are ideal for aerospace and defense, rechargeable batteries for electric vehicles, and quantum technologies as well.

A major application of nickel in quantum technologies is in iron-77~80% nickel alloy used for magnetic shielding of superconducting quantum computers. An emerging application is the substitution of current cryostat dilution refrigerators by synthesizing high-performance magnetocaloric materials with holmium (Ho).[59]

U.S. net import reliance on nickel in 2025 is 41%, which on its face suggests lower exposure to supply disruption than for niobium. [60] Ownership structure, however, complicates this read. Indonesia, home to the world's largest nickel reserves and a rapidly growing refining sector, has dominated the production of nickel, accounting for nearly 66% of global output in 2025.[61]

However, much of this capacity is foreign owned. Chinese companies or shareholders ultimately control at least 75% of Indonesia's nickel refining capacity, but this is hidden behind layers of shell companies that mask foreign ownership. U.S. and European Union automakers have only recently entered the industry, often by partnering with or sourcing from established Chinese companies, even as their governments grow increasingly cautious of Chinese dominance in critical supply chains.

---

[58] E. I. Malevannaya et al., An engineering guide to superconducting quantum circuit shielding, Appl. Phys. Rev. 12, 031334 (2025), *supra* note 56.
[59] Y Zhnag et al., Exploration of rare-earth cobalt nickel-based magnetocaloric materials for hydrogen liquefaction, J. Mater. Sci. Tech., 159, 163 (2023), https://doi.org/10.1016/j.jmst.2023.04.001.
[60] U.S. Geological Survey. Mineral Commodity Summaries 2026, *supra* note 18.
[61] U.S. Geological Survey. Mineral Commodity Summaries, Nickel 2026, https://pubs.usgs.gov/periodicals/mcs2026/mcs2026-nickel.pdf

Looking ahead, nineteen nickel refineries accounted for over 90% of Indonesian nickel production capacity, as measured by capacity rather than actual output, in 2023. Indonesia's eight million metric ton nickel refining capacity (as of 2023) initially appears to be distributed across thirty-three companies. However, ownership tracing reveals that Chinese companies or Chinese shareholders control 61% of the refining capacity, while Indonesian companies or shareholders control just 13%. Further tracing uncovers that Chinese beneficial ownership exceeds 75%. As Indonesia aims to use the nickel industry for economic growth, this substantial foreign influence could limit its ability to control and shape the industry for its benefit. As a result of this strategic movement by China, two Chinese companies have considerable shares in refineries that account for over 70% of Indonesia's nickel refining capacity: Tsingshan Holding Group and Jiangsu Delong Nickel Industry Co Ltd.[62], [63]

It will be challenging for other countries to meet domestic demand without access to Indonesian nickel. China's entrenched position in the industry makes it difficult for companies to establish a fully independent and secure supply chain outside of China. The reliance on Chinese-controlled nickel production not only raises concerns about supply-chain resilience, but also places U.S. and European manufacturers amid increasingly restrictive policies on trade with China.

In May 2025, the U.S. Department of Defense, under the Defense Production Act, Title III, awarded a grant of $7 million to a Missouri company to develop a hydrometallurgical demonstration plant to produce cobalt and nickel products. The plant would be capable of extracting the metals from a variety of feedstocks.[64]

Even though the DoD's activity is timely, the United States is currently in the preliminary stages of investing in exploration and mining, which is the easiest and most intuitive part of rebuilding the supply chain. The real challenge in rebuilding the U.S. rare-earth supply chain lies in acquiring advanced smelting and refining technologies and expanding production capacity. These are fundamental issues that cannot be resolved with simple or superficial measures. Furthermore, given the case of Lynas's Malaysian smelting facility parts supply interruption following the U.S.-Australia rare-earths alliance in November 21, 2025,[65] even if rare-earth production facilities were to be operational, most consumables and equipment related to their operation, maintenance, and repair are controlled by Chinese companies. Therefore, even if full-scale production facilities were established, completing the supply chain would not be an easy task.

---

[62] The Center for Advanced Defense Studies (C4ADS) commentary, Refining Powder, Feb. 4, 2025, https://c4ads.org/commentary/refining-power/.
[63] Mining Technology, China's control over 75% of Indonesia's nickel capacity raises concerns, Feb. 6, 2025, https://www.mining-technology.com/news/chinas-indonesias-nickel-capacity/.
[64] DOD Awards $7 Million to Enhance Domestic Nickel and Cobalt Supply Chains, (Mar. 15, 2024), https://www.war.gov/News/Releases/Release/Article/3708859/dod-awards-7-million-to-enhance-domestic-nickel-and-cobalt-supply-chains/.
[65] Washington Post Malaysia, supra note 9.

| Element | Type | Production amount by 1st producer, 2025 (metric tons) | Sum of 1st and 2nd production, 2025 (metric tons) | Wold production, 2025 (metric tons) | Fraction of 1st producer (%) | Fraction of 1st and 2nd producer (%) | 1st Producer | 2nd producer |
|---|---|---|---|---|---|---|---|---|
| Abrasives | Al2O3 | 800000 | 890000 | 1300000 | 61.54 | 68.46 | China | Austria |
| | SiC | 450000 | 530000 | 1000000 | 45.00 | 53.00 | China | Norway |
| Al | | 45000 | 49200 | 74000 | 60.81 | 66.49 | China | India |
| Sb | | 40000 | 72000 | 110000 | 36.36 | 65.45 | China | Russia |
| As | | 30000 | 54000 | 61000 | 49.18 | 88.52 | Peru | China |
| Asbestos | | 310000 | 560000 | 960000 | 32.29 | 58.33 | Russia | China |
| Barite | | 3000 | 5200 | 8700 | 34.48 | 59.77 | India | China |
| Bauxite | | 150000 | 247000 | 440000 | 34.09 | 56.14 | Guinea | Australia |
| Be | | 230 | 310 | 430 | 53.49 | 72.09 | US | Brazil |
| Bi | | 14000 | 15000 | 16000 | 87.50 | 93.75 | China | Korea |
| B | Borate | 1500 | 1880 | 2920 | 51.37 | 64.38 | Turkey | Bolivia |
| Bromine | | 200000 | 310000 | 430000 | 46.51 | 72.09 | Israel | Jordan |
| Cd | | 9500 | 13800 | 23000 | 41.30 | 60.00 | China | Korea |
| Cr | | 23000 | 32000 | 51000 | 45.10 | 62.75 | South Africa | Turkey |
| Co | | 230000 | 274000 | 310000 | 74.19 | 88.39 | Congo | Indonesia |
| Cu | refinery production | 14000 | 16800 | 29000 | 48.28 | 57.93 | China | Congo |
| Diamond | | 16 | 23 | 38 | 42.11 | 60.53 | Russia | Congo |
| Fluospar | | 6000 | 7500 | 10000 | 60.00 | 75.00 | China | Mexico |
| Ga | Primary | 900000 | 906000 | 906000 | 99.34 | 100.00 | China | Russia |
| | Production capacity | 1600000 | 1601000 | 1700000 | 94.12 | 94.18 | China | Russia |
| Garnet | | 350000 | 630000 | 730000 | 47.95 | 86.30 | Australia | China |
| Ge | kg | 36656 | 36656 | 36656 | 100.00 | 100.00 | China | - |
| Graphite | | 1400000 | 1480000 | 1800000 | 77.78 | 82.22 | China | Madagascar |
| He | Gas | 81 | 144 | 190 | 42.63 | 75.79 | US | Qatar |
| In | | 760 | 940 | 1100 | 69.09 | 85.45 | China | Korea |
| Iodine | | 23000 | 32000 | 34000 | 67.65 | 94.12 | Chile | Japan |
| Iron and Steel | Pig iron | 830 | 928 | 1300 | 63.85 | 71.38 | China | India |
| | Raw steel | 980 | 1140 | 1900 | 51.58 | 60.00 | China | India |
| | Ore | 600000 | 860000 | 1600000 | 37.50 | 53.75 | Australia | Brazil |
| Pb | | 1900 | 2380 | 4500 | 42.22 | 52.89 | China | Australia |
| Li | | 92000 | 154000 | 290000 | 31.72 | 53.10 | Australia | China |
| Mg | metal | 950 | 1010 | 1100 | 86.36 | 91.82 | China | Russia |
| Mn | | 7600 | 12600 | 20000 | 38.00 | 63.00 | South Africa | Gabon |
| Hg | | 200 | 205 | 210 | 95.24 | 97.62 | China | Kyrgyzstan |
| Mo | | 97000 | 139000 | 260000 | 37.31 | 53.46 | China | Chile |
| Ni | | 2600000 | 2870000 | 3900000 | 66.67 | 73.59 | Indonesia | Philippines |
| Nb | | 104000 | 110000 | 112000 | 92.86 | 98.21 | Brazil | Canada |
| N | Ammonia | 49000 | 64000 | 160000 | 30.63 | 40.00 | China | India, Russia |
| P | | 110000 | 146000 | 250000 | 44.00 | 58.40 | China | Morocco |
| Pt | | 120000 | 140000 | 170000 | 70.59 | 82.35 | South Africa | Russia |
| Pd | | 84000 | 154000 | 190000 | 44.21 | 81.05 | Russia | South Africa |
| REE | mine production | 270000 | 321000 | 390000 | 69.23 | 82.31 | China | US |
| Re | | 30000 | 50000 | 81000 | 37.04 | 61.73 | Chile | China |
| Se | | 2000 | 2640 | 3800 | 52.63 | 69.47 | China | Japan |
| Sc | | 80 | 80 | 90 | 88.89 | 88.89 | China | - |
| Si | metal | 4000 | 4180 | 4600 | 86.96 | 90.87 | China | Brazil |
| Ag | | 6300 | 9900 | 26000 | 24.23 | 38.08 | Mexico | Peru |
| Sr | | 250000 | 350000 | 450000 | 55.56 | 77.78 | China | Spain |
| Ta | | 1300 | 1700 | 2500 | 52.00 | 68.00 | Congo | Rwanda |
| Te | | 800 | 867 | 1000 | 80.00 | 86.70 | China | Russia |
| Sn | | 71000 | 132000 | 290000 | 24.48 | 45.52 | China | Indonesia |
| Ti | sponge | 260000 | 313000 | 370000 | 70.27 | 84.59 | China | Japan |
| W | | 67000 | 70000 | 85000 | 78.82 | 82.35 | China | Vietnam |
| V | | 82000 | 103000 | 110000 | 74.55 | 93.64 | China | Russia |
| Zn | | 4100 | 5600 | 13000 | 31.54 | 43.08 | China | Peru |
| Zr | | 400 | 670 | 1200 | 33.33 | 55.83 | China | South Africa |
| | | | | | | | | |
| 28 | 1st producer >50% | | | | | | | |
| 7 | 2nd largest producer under threatened merging by China | | | | | | | |
| 4 | Under controlled by China | | | | | | | |
| 12 | Producing more than 70% by China | | | | | | | |
| 6 | Producing more than 50% by Allies | | | | | | | |

*Table 3. Critical Level I materials (analysis based on the USGS 2026 commodity summaries).*[66]

[66] U.S. Geological Survey. Mineral Commodity Summaries 2026, *supra* note 18.

This data shows elements for which the top producing country accounts for more than 50% of global production, according to the 2026 U.S. Geological Survey Mineral Commodity Summaries.[67] Elements highlighted in yellow are those for which the top producing country accounts for an overwhelmingly larger portion than other producing countries, while elements highlighted in orange are those for which China and countries within China's reach are the second-largest producers. Including these two countries, these elements account for an overwhelmingly large portion of global production. Elements highlighted in pink, while not shown in the statistics, are those for which a single largest producer accounts for more than 60% of global production, effectively leaving production vulnerable to Chinese control or influence. The most problematic elements here are the pink elements (Sb, Co, Ni, Nb), whose production is almost entirely monopolized by a single country and is realistically subject to Chinese influence or control, while the orange elements (As, Cd, In, Mo, Re, Sr, Ta) are those for which the second-largest producer is either under threat or being acquired by China.

### 4. Use Case 2: QT in Harsh Environments (Space SNSPDs; Aerospace and Arctic Missions)

The modern space sector—encompassing satellite communications, Earth observation, deep-space exploration, and launch-vehicle innovation—is increasingly dependent on a diverse range of critical minerals. As states and private firms invest in space as both a strategic domain and a commercial frontier, demand for specialized materials has grown in parallel.[68] From lightweight alloys and thermal-resistant composites to rare elements essential for sensors, optics, propulsion, and electronics, critical minerals underpin every stage of space technology. In an era of intensifying geopolitical competition, shifting energy policies, and mounting supply-chain pressure, securing access to these minerals has become a pivotal concern for governments and industry alike. From rare-earth magnets and radiation-hardened semiconductors to high-performance alloys and propellants, the materials enabling satellite communications, space exploration, and orbital infrastructure are becoming central to national critical-minerals strategies.[69]

Governments are correspondingly placing greater emphasis on supply-chain transparency and resilience to ensure uninterrupted access to essential elements. Disruptions—whether driven by geopolitical tensions, trade restrictions, or limited global production—can threaten satellite networks, navigation systems, and advanced space missions. Bismuth, phosphorus, lithium, and germanium, for example, form part of the material backbone of modern space optics, enabling spacecraft to operate under conditions in which conventional materials would fail.

[67] *Ibid.*
[68] Critical Minerals and Space Technologies, SFA Oxford, https://www.sfa-oxford.com/knowledge-and-insights/critical-minerals-in-low-carbon-and-future-technologies/critical-minerals-for-space-and-technology/.
[69] *Ibid.*

Their distinctive properties, ranging from infrared transparency to radiation resistance, are important for remote sensing, imaging, high-speed communications, and laser-based systems.[70]

The space environment presents extreme challenges for quantum systems, including severe temperature fluctuations, ionizing radiation, vacuum exposure, vibration, and the need for long-term operational stability. Materials such as bismuth, phosphorus, lithium, and germanium are relevant to these demands through their integration into germanium IR detectors, bismuth-tellurite fiber optics, and lithium niobate ($LiNbO_3$) modulators.[71] The central point for this study is that harsh-environment performance is not merely a packaging problem; it is also a materials and supply-chain problem.

The second use case therefore widens the analysis from supply concentration to environmental survivability. In Arctic, aerospace, and space deployments, a component can be strategically critical not only because it is difficult to source, but because its degradation under radiation, thermal cycling, shock, or vacuum can silently erase the security value of the system in which it is embedded. Recent defense-oriented quantum analyses increasingly treat clocks, sensors, secure communications payloads, and resilient position-navigation-timing functions as operational enablers rather than laboratory curiosities.[72]

Different types of superconducting single-photon detectors (SPDs) exist, with differing operating principles, device structures, and material systems. The superconducting nanowire single-photon detector (SNSPD) is a quantum-limit superconducting optical detector based on the Cooper-pair-breaking effect induced by a single photon. Compared with alternative detector classes, SNSPDs offer higher detection efficiency, lower dark-count rates, higher counting rates, and lower timing jitter, and they have therefore been widely applied in quantum information processing, including quantum key distribution (QKD) and optical quantum computation. [73]

Over the past two decades, many superconducting materials have been explored for SNSPD fabrication, including Nb, NbN, NbTiN, NbSi, WSi, MoSi, MoGe, MoN, TiN, and $MgB_2$. Among these, Nb(Ti)N, WSi, and MoSi have emerged as leading candidates, achieving system detection efficiencies above 90%.[74] Commercialization efforts are also underway. At the time of writing, six companies—ID Quantique (Switzerland), CNPHOTEC (China), Photon Spot (USA), Quantum Opus (USA), SCONTEL (Russia), and Single Quantum (Netherlands)—were working on SNSPD commercialization.[75]

---

[70] *Ibid.*
[71] *Ibid.*
[72] Michal Krelina, *The Prospect of Quantum Technologies in Space for Defence and Security*, 65 Space Pol'y 101563 (2023), https://www.sciencedirect.com/science/article/pii/S0265964623000255.
[73] L. You, Nanophotonics 9, 2673 (2020), Superconducting nanowire single-photon detectors for quantum information, https://doi.org/10.1515/nanoph-2020-0186.
[74] *Ibid.*
[75] *Ibid.*

| | Element | Application condition | Fabrication method | Specification | Product type |
|---|---|---|---|---|---|
| **Materials** | InP | Avalanch Photon Diode (APD) | | 1550nm wave length | InP wafer |
| | | Epitaxial Layer unifomity | | | |
| | | Dislocation density | | | |
| | | Epitaxial growth surface condition | | | |
| | | Lattice parameter | | | |
| | | Defect minimization | | | |
| | InGaAs | Doping thickenss control | | | Epitaxial wafer |
| | | Dopant concentration | | | |
| | | | Zn-diffusion method | | |
| | | | Gas-phase diffusion | | |
| | | | Solid-phase diffusion | | |
| | Signal-to Noise ration (SNR) | | | | ZnP2 |
| | Operation mode | | | | |
| | | Geiger mode | | | |
| | | Gate mode | | | |
| | SiO2 | Core | | | SiO2 PLC wafer |
| | | Clad | | | |
| | | | Synthesize method | | |
| | | | Chemical vapor deposition (CVD) | | Si wafer |
| | | | Flame hydrolysis deposition (FHD) | | |
| | | | | | |
| | | | | | |
| | | | | | |
| | | | | | |
| | LiNbO3 | Thin film thickness | | <1micron or <2micron | LNOI wafer |
| | | bulk preparation | | | |
| | SNSPD | Crystalline | | <100nm | |
| | | NbN | | | |
| | | NbTiN | | | SNSPD superconductor |
| | | Amorphous | | | |
| | | WSi | | | |
| | | MoSi | | | |
| | **Element** | **Application condition** | **Fabrication method** | **Specification** | **Product type** |
| **Component** | Single mode fiber (SMF) | | | | SMF fiber |
| | | GeO2 coating | | | |
| | | Core diameter | | <10micron | |
| | | Cladd diameter | | 125micron | |
| | | Scattering loss | | <0.2dB/km@1.55micron dia. | |
| | | Absorption loss | | | |
| | SMF Preform | | | | |
| | | Drawing tower | | | |
| | | Length | | 60-70cm | |
| | | Diameter | | 20-25mm | |
| | | | Sythesize method | | |
| | | | Modified chemical vapor deposition (MCVD) | | |
| | | | Vapor phase axial deposition (AD) | | |
| | | | Outside vapor phase oxidation (OVPO) | | |
| | | | Plasma activated chamical vapor deposition (PCVD) | | |
| | Laser diode | | | | Laser diode |
| | | FP-LD | | | |
| | | p-n junction | Distributed feedback (DFB) | | |
| | | | Distributed Bragg reflector (DBR) | | |
| | | Distributed feed back type (DFB-LD) | | | |
| | | InGaN/GaN, SiC | | (active layer/wafer) | |
| | | AlGaInP/GaAs | | wireless communication (visible light wave) | |
| | | AlGaAs/GaAs | | | |
| | | InGaAs/GaAs | | | |
| | | AlGaAsP/InP | | wire comminucation (1550-1310nm wave) | |
| | | AlGaAsSb/GaAsSb | | | |
| | | Output power | | high power | |
| | | Active wavelength | | single wavelength | |
| | | Range | | <1MHz | |
| | | Relative noise | | ext. low level | |
| | APD | | | | APD |
| | | Breakdown voltage | | | |
| | | III-V compound | | | |
| | | InP | | | |
| | | InGaAs/InP | | Separation Absorption and Multiplication (SAM) structure | |
| | | InGaAsP | | Separated Absorption Grading and Multiplication (SAGM) structure | |
| | | GaSb | | | |
| | | AlGaSb | | defect problem to growth | |
| | | II-VI compound | | | |
| | | HgCdTe | | CdTe(Eg=1.5eV) | |
| | | | | lattice defect problem | |
| | | Ge | | | |
| | | n-type doping | Field contorl layer | | |
| | Pahse modulator | | | | Phase Modulator |
| | | Pockets cell | | | |
| | | Potassium Di-deuterium Phosphate (KDP-DKDP) | | | |
| | | Potassium Titanyl Phosphate (KTP) | | | |
| | | beta-Barium Borate (BBO) | | | |
| | | Lithiium Niobate (LiNbO3) | | | |
| | | Lithium Tantalate (LiTaO3) | | | |
| | | Ammonium Dihydrogen Phosphate (NH4H2PO4) | | | |
| | | Cadmium Telluride (CdTe) | | | |
| | Intensity modulator | | | | Intensity Modulator |
| | | Acousto-optic | | | |
| | | Diffractor | | | |
| | | Electro-optic | | | |
| | | Polarizer | | | |
| | | Mach-Zehnder (MZ) | | | |
| | | Spatial Light Modulator | | | |
| | | Electro-Absoprtion | | | |
| | | | Franz-Keldysh Effect | | |
| | Anti-Photon Bumber Splitting (PNS) | | | | |
| | | Decoy pulse | | | |
| | Optical Attenuator | | | | Optical Attenuator |
| | | Variable optical attenuator (VOA) | | | |
| | | | | Bit-Error- rate (BER) | |
| | | | | Air gap | |
| | | | | Misaligned splicing | |
| | | | | Optical fiber bending | |
| | | | | Beam Collimating | |
| | Interferometer | | | | Interferometer |
| | | SiN/SiO2 | | 0.2mm, 1ps | |
| | | Faraday Rotator mirror (FRM) | | | |
| | | Thermo Electric cooler (TEC) | | | |
| | | Faraday-Michelson system | | <2.8% QBER | |
| | | Power Line Communication (PLC) | | | |
| | | Multi-Mode Interference (MMI) | | width 13micron, length 313micron, 50% | |
| | | Wavelength Independent coupler (WINC) | | 8 degree | |
| | | Thermistor | | | |
| | Qauntum Random Number Generator (QRNG) | | | | QRNG module |
| | | | | | |

*Table 4. Major materials and components for QKD of quantum communication.*

Today's leading SNSPDs for telecom wavelengths (~1550 nm), often based on NbN, WSi, or MoSi on silicon or related substrates, deliver excellent performance in laboratory conditions, including very high detection efficiency and low timing jitter. Although at least one private firm has attempted to adapt QKD systems for space use, no commercially fielded space-qualified product has yet emerged, and existing devices are not optimized for space deployment.[76] They remain sensitive to magnetic fields, while radiation can generate lattice defects or trapped charges in the superconducting film or substrate, increasing dark counts, reducing efficiency, or introducing transient failure modes. In adjacent sectors, such as satellite CMOS sensing, radiation-hardening techniques include shielding, redundancy, and specialized materials. For SNSPDs, preliminary evidence suggests that amorphous superconductors such as WSi may be more resistant to radiation-induced defects because they lack a crystalline lattice, but this remains insufficiently studied. In short, existing SNSPD designs perform superbly under controlled conditions, but no current design is yet proven across the radiation and thermal extremes of space, and available data on long-duration aging under such conditions remain limited.

This gives rise to a mission-assurance problem as much as a device-physics problem. We need to evaluate how radiation, temperature cycling, vibration, and electromagnetic interference affect SNSPDs and related photonic components, and then develop design improvements that minimize these effects. The key unknowns are practical and operational: How much does radiation increase dark counts or reduce detection efficiency over time? Are the dominant failure modes gradual and monitorable, or sudden and mission-ending? Can detectors be designed to fail gracefully, reconfigure safely, or self-correct? Without answers to these questions, any QKD payload relying on current detector architectures may have a severely limited operational lifetime in orbit—potentially measured in weeks or months rather than years.

Recent work on quantum-security threat discovery strengthens this argument by showing that stakeholders often identify socio-technical risks that purely technical reviews overlook: governance gaps, human-factor vulnerabilities, hidden dependencies, and broader societal impacts.[77] For harsh-environment QT, "fails gracefully" should therefore be understood as both a device-level and governance-level requirement.

For mission assurance, a central question is whether key components fail gracefully, that is, whether performance degrades in a predictable and monitorable manner that permits safe fallback, recalibration, or mission reconfiguration without silent failure modes. This concern can be framed in system terms as follows:

---

[76] Qubitrium, https://qubitrium.tech (last visited Apr. 20, 2026).

[77] Steven Umbrello, Pieter E. Vermaas, Indika Kumara, Joost Alleblas, Stefan Driessen & Willem-Jan van den Heuvel, *Quantum Security Threat Discovery: A Value Sensitive Design Approach to Discovering Security Risks of Quantum Sensing at the Port of Moerdijk*, 19 NanoEthics art. 8 (2025), https://doi.org/10.1007/s11569-025-00475-y.

| Mission stressor | Component / dependency | Primary metric(s) | System-level implications (example) |
|---|---|---|---|
| Radiation (ionizing / displacement) | SNSPD materials/substrates; readout electronics | Dark count rate; detection efficiency; timing jitter | Higher QBER or reduced link margin in QKD; reduced fidelity in photon-based experiments |
| Thermal cycling | Packaging; interfaces; cryostat/thermal control | Stability / drift; failure rate | Calibration burden; intermittent downtime; reduced mission availability |
| Vibration / shock | Optical alignment; fiber coupling; mechanical supports | Coupling loss; alignment tolerance | Link loss and intermittent operation; increased need for redundancy |
| EM interference | Readout chain; shielding materials | Noise floor; error rate | Higher classical-processing overhead; reduced robustness under field conditions |

*Table 5: Mission stressor–to–metric mapping for space-qualified SNSPD systems.*

One possible route beyond the current state of the art is the development of radiation-hardened single-photon detector systems. Promising approaches include amorphous superconducting nanowire materials, which may be less vulnerable to lattice-disruption effects; nanowire geometries with redundant paths or self-healing layouts so that local defects do not disable the entire detector; and additional protective elements such as on-chip shielding layers or fault-tolerant readout circuits. An active calibration mechanism—for example, an internal LED that periodically tests detector response and recalibrates bias settings to compensate for drift—may also be valuable. The goal would be a detector that exhibits only limited dark-count inflation under radiation exposure and retains most of its efficiency after prolonged operation in harsh environments. If achieved, such performance would set a new benchmark for quantum hardware suitable for space and other radiation-intensive settings.

The broader point extends beyond SNSPDs. Superconducting and photonic qubits rely on fragile quantum states sustained by superconducting or photonic materials. In superconducting systems, paired electrons (Cooper pairs) flow without resistance and help preserve the qubit's superposition state. If the circuit is heated or otherwise perturbed, these pairs can break into quasiparticles, causing decoherence and limiting performance. Relevant sources of decoherence include cosmic rays, gamma radiation, fluctuating magnetic and electric fields, thermal energy, and crosstalk between qubits.[78]

Cosmic rays and background gamma radiation are especially important because they can produce large, correlated errors in superconducting qubits. Cosmic rays originate from high-energy particles entering the atmosphere and producing muons that penetrate deeply into matter at the Earth's surface. When such particles traverse a quantum chip, they can deposit substantial energy in the processor substrate, transiently heating the chip. Gamma rays from natural background sources can do likewise. These pulsed energy events matter because correlated qubit errors in time and space can undermine error correction by orders of magnitude.[79] In space and polar environments, where radiation exposure, thermal swings, and operational constraints are amplified, these effects become even more salient.

[78] MIT News, Cosmic rays may soon stymie quantum computing, Aug. 26 (2020), https://news.mit.edu/2020/cosmic-rays-limit-quantum-computing-0826.

The harsh environment of space also places pressure on the underlying materials platform more broadly. Germanium, gallium-based compounds (i.e., indium-gallium-zinc-oxide: IGZO), phosphorus, antimony, lithium, and arsenic derivatives underpin spacecraft electronics, solar panels, sensors, and communication systems, while their electronic and optical properties support durable operation over long mission durations.[80] Policies such as the U.S. CHIPS Act, Europe's Horizon Europe program, and Japan's Moon-to-Mars roadmap therefore matter not only for semiconductors in general, but also for gallium nitride (GaN), gallium arsenide (GaAs), and related materials increasingly relevant to space electronics.[81]

Alongside this trend, DARPA also launched the Robust Quantum Sensors (RoQS) program to advance QS technology by developing sensors that resist performance degradation under real-world conditions and are suitable for deployment on Department of Defense platforms. While quantum sensors have demonstrated exceptional capabilities in controlled settings, their performance often declines in dynamic environments due to factors such as vibration and electromagnetic interference. The RoQS program is designed to address this fragility at its source. Rather than relying on bulky shielding or isolation techniques that are impractical for broad deployment, it explores new sensor designs and architectures capable of suppressing environmental disturbances.[82]

More speculatively but importantly, the interface of QS and quantum materials may open new pathways for harsh-environment systems. Novel many-body states in topological superconductors, quantum-critical materials, and frustrated magnets may enable new modes of sensing, while local quantum sensors such as color centers in diamond may in turn illuminate the properties of these materials. Likewise, photonic microcavities based on highly nonlinear materials, including III-V semiconductors and ferroelectrics, may support quantum frequency combs, chip-scale entangled-photon sources, and dynamic quantum meta-surfaces relevant to secure satellite communications and QKD. Scientists are also exploring single-photon emitters and detectors in amorphous II-VI materials as a potential platform for scalable quantum photonics in secure space communications. New materials in this area should be evaluated not only for raw device performance, but also for technical feasibility, integration potential, and compatibility with present and future quantum-secured communications architectures.

---

[79] J. Martinis, npj Quantum Information 7, 90 (2021), Saving superconducting quantum processors from decay and correlated errors generated by gamma and cosmic rays, https://doi.org/10.1038/s41534-021-00431-0

[80] Park et al., Mater. Sci. Semiconductor Processing, 204, p. 110320 (2026), https://doi.org/10.1016/j.mssp.2025.110320/; and Z. Muhammad et al., Adv. Mater. Tech. 8, p. 2200539 (2023), https://doi.org/10.1002/admt.202200539.

[81] SFA Oxford.

[82] Defense Advanced Research Projects Agency (DARPA), The Robust Quantum Sensors (RoQS) program, https://www.darpa.mil/research/programs/roqs-robust-quantum-sensors.

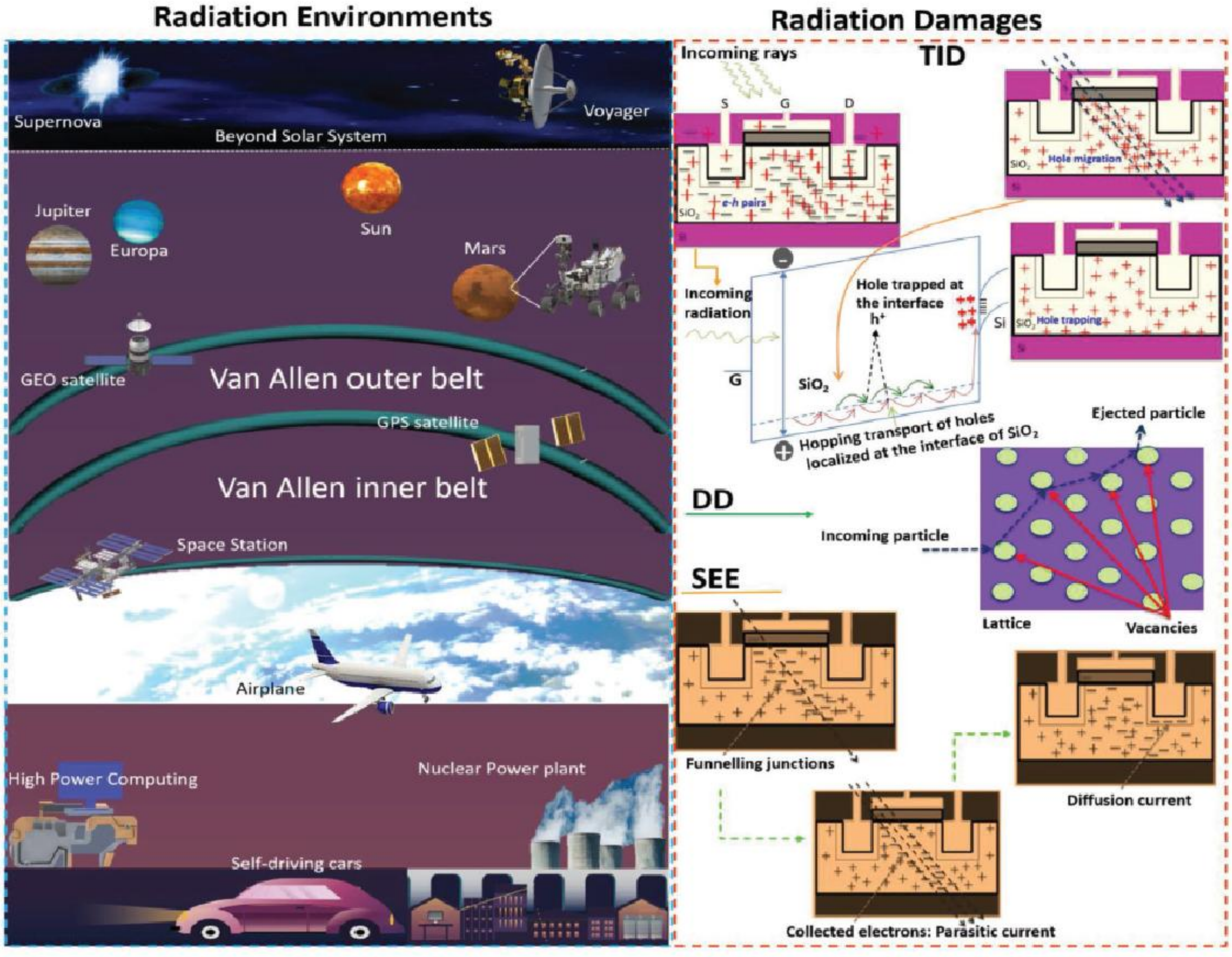

*Figure 5. High energy radiation environments and radiation damage semiconductor devices.*[83]

[83] Z. Muhammad et al., Adv. Mater. Tech. 8, p. 2200539 (2023), https://doi.org/10.1002/admt.202200539.

| Radiation Tolerant materials [dose $cm^{-2}$] | Radiation properties at various irradiation dose [ions $cm^{-2}$] | | | | |
|---|---|---|---|---|---|
| | $V_{th}$ shift [V] | Mobility [$cm^2 V^{-1} s^{-1}$] | $I_{ON}/I_{OFF}$ | Change in SS [V $dec^{-1}$] | Refs. |
| Si (at $10^{15}$) | =0.8 | increase by 5% | decreased by =0.1 | =1 | [118] |
| ZnO (at $10^{14}$) | −15 | decreased by 1.5 | No change | 1.02 | [29] |
| $In_2O_3$ (at $10^{12}$) | =0.1 | Degraded | decrease by $10^1$ | 0.75 | [35] |
| $Ga_2O_3$ (at $10^{15}$) | 1.2 | decreased by 2.02 | decreased by 4.7 | =0.06 | [12] |
| IZO (at $3 \times 10^{11}$) | −5.7 | increased by 0.16 | decrease by $10^1$ | 0.7 | [31] |
| IGZO (at $10^{15}$) | −88 | increased by 4.4 | decreased by $10^5$ | 55 | [22] |
| $SnO_2$ (at $10^{12}$) | −3 | No change | No change | 3.5 | [34] |
| SnO (at $10^{14}$) | ≤0.4 | ≤0.02 | decreased by =$10^4$ | =0.2 | [119] |
| ZTO (at $10^{15}$) | 2.3 | increased by 3 | decreased by $10^3$ | 7.4 | [13] |
| GaN (at $10^{18}$) | =1 | increased by=1.8 | decreased by $10^2$ | =6.7 | [120] |
| GaAs (at $10^{15}$) | 0.4 | decreased by=1.5 | No change | – | [121] |
| 4H-SiC (at $10^{13}$) | −2.8 | increased by 1.2 | increased by=120 | =2.3 | [122] |
| $WS_2$ (at $10^{15}$) | <0.1 | degraded | decreased by 6350 | – | [50] |
| $MoS_2$ (at $10^{16}$) | −2 | decreased by 3.5 | degraded | 2.85 | [36] |
| $MoSe_2$ (at $6 \times 10^{10}$) | +8 | increased twice | increased by=7.4 | =0.2 | [123] |
| $WSe_2$ (at $10^{15}$) | =8 | Same | decreased by=0.6 | =0.1 | [124] |
| $Cs_2CrI_6$ (at $10^{16}$) | =40 | No change | – | – | [83] |

*Figure 6. Radiation immune properties of various WBGs (wide band gaps).*[84]

### 4.1. Shielding and Radiation-Hardening Materials for Harsh Environments

Recently, superconducting multi-qubit processors have demonstrated quantum supremacy and quantum error correction below the surface-code threshold. Yet decoherence-induced loss of quantum information from environmental noise remains a major challenge. Considerable work has focused on suppressing external noise sources such as flux noise, charge noise, the Purcell effect, and stray electromagnetic modes in cryogenic setups. By contrast, protection against broadband spontaneous radiation—from surrounding equipment, infrared radiation, cosmic rays, and background radiation—remains insufficiently solved. Depending on the propagation path of stray radiation, whether through free space or coaxial microwave transmission lines, different shielding and filtering strategies are required.[85]

External radiation can break Cooper pairs and generate quasiparticles, which, when tunneling through a Josephson junction, contribute both to qubit energy relaxation and to dephasing. Variations in external magnetic fields and induced charges near the Josephson-junction region can also produce uncontrolled changes in Hamiltonian parameters, including the Josephson energy, the superconducting phase difference, and the qubit frequency.[86]

This is where supporting materials become important. In harsh environments, shielding materials cannot be assessed solely on classical engineering grounds such as mass, cost, or manufacturability. They must also be evaluated for their compatibility with quantum performance requirements, including magnetic shielding, radiation attenuation, cryogenic behavior, and long-duration stability. In this context, materials such as tantalum, tungsten, vanadium alloys, and related advanced coatings may merit closer examination, particularly

[84] *Ibid.*
[85] Malevannaya, *supra* note 56.
[86] G. Ithier et al., Decoherence in a superconducting quantum bit circuit, Phys. Rev. B 72, 134519 (2005).

where they can improve radiation tolerance or reduce unwanted thermal and electromagnetic coupling.

More broadly, future radiation shielding for quantum and space applications may need to move beyond traditional shielding materials, which are often heavy, toxic, or expensive. Emerging approaches such as cold-spray or kinetic-fusion coatings—for example, refractory metal based cold-spray platform—may offer lightweight, durable, and potentially radiation-resistant protective layers with low heat input and minimal waste.[87] Whether such approaches are suitable for quantum-relevant components remains an open question, but they illustrate the wider design space for materials-based mitigation from nuclear infrastructure to space missions.

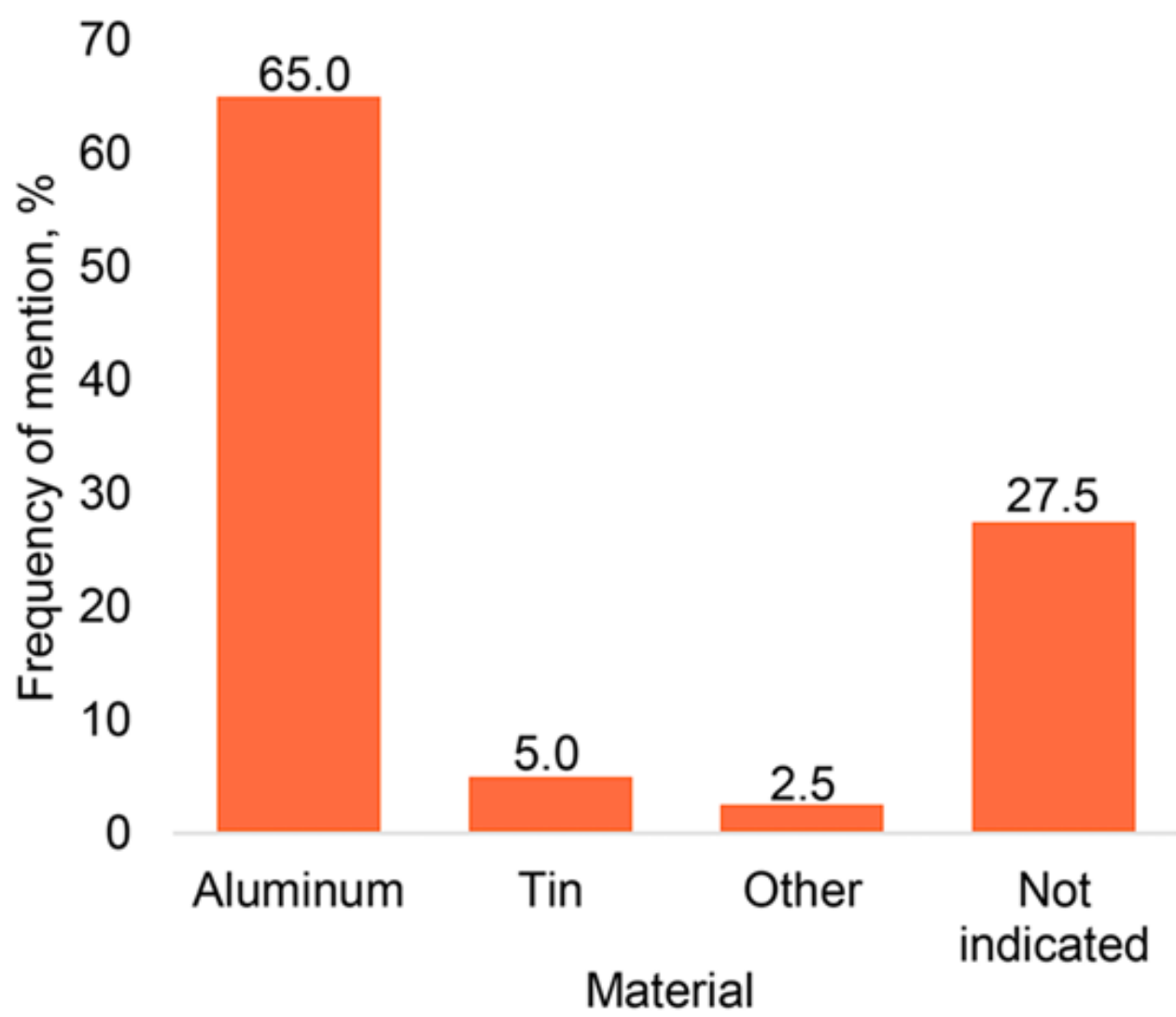

*Figure 7. Typical materials used for superconducting shields.*[88]

## 5. Security and Operational Deployment: From Materials Criticality to Quantum-Secure Communications

The preceding sections identify the upstream CMM dependencies and the harsh-environment failure modes that can degrade quantum performance in mission settings. This section translates those dependencies into security and deployment consequences, focusing on how component fragility—e.g., detectors, optics, cryogenics, and radiation-sensitive electronics—propagates into system-level security outcomes. We begin with quantum-secure communications—PQC migration and QKD/quantum networking—because their operational viability is tightly coupled to the same mission stressors analyzed above, and because the relevant security objective is continuity of security rather than intermittent "quantum-secure" operation with insecure fallback. We then extend the same mission-assurance logic to adjacent security-relevant domains, including QC-enabled security applications, QS and networking, and quantum-AI hybrids.

[87] Titomic to provide radiation shielding for Fleet Space's Alpha satellite, June 10 (2022), https://titomic.com/2022/06/10/titomic-to-provide-radiation-shielding-for-fleet-spaces-alpha-satellite/.

[88] E. I. Malevannaya et al., An engineering guide to superconducting quantum circuit shielding, Appl. Phys. Rev. 12, 031334 (2025), *supra* note 56.

Section 5 connects materials criticality to a broader control plane of standards, certification, procurement, export controls, and operational fallback. The core claim is that secure deployment requires aligning the physical front end of quantum supply chains with the logical front end of cryptographic transition. If the former fails, mission-capable hardware does not scale; if the latter fails, hardware may operate atop insecure networks, legacy trust anchors, or brittle fallback protocols.

### 5.1. Quantum-Secure Communications: PQC, QKD, and Quantum Networking

From an operational perspective, the security objective is continuity of security: quantum-secure modes should not degrade into insecure fallback under stress, such as link loss, degraded detector performance, or subsystem failure. In this manuscript, the harsh-environment mission profile is therefore mapped to measurable device and network metrics that determine whether secure operation can be maintained, degraded safely, or must be suspended.

Quantum-secure communication has become a global strategic priority, driven by two parallel trends: the rise of QC, which threatens traditional cryptography, and advances in quantum cryptography, which offer new methods for secure communication.[89] Over the next decade, large-scale quantum computers may break current encryption methods such as RSA and ECC. This creates an urgent harvest-now-decrypt-later threat: adversaries may intercept sensitive encrypted data today and store it for decryption once sufficiently capable quantum computers become available. Experts accordingly warn that meaningful quantum decryption capabilities could emerge as early as the 2030s.

In February 2026, European security officials reported unusual movement by two Russian space satellites that they believed may have intercepted the communications of at least a dozen key satellites over the continent.[90] Officials reportedly feared not only compromise of sensitive information transmitted by the satellites, but also possible interference with satellite control, trajectory, or safety. While China and the United States have developed similar technologies, Russia has one of the most advanced space-spying programs and has been more aggressive in using satellites to stalk those of other states. The episode also highlights a structural vulnerability: sensitive information—notably command data for some European satellites—may remain insufficiently encrypted because most systems were launched years ago without advanced onboard computing or robust encryption capabilities. In that sense, satellite networks can become an Achilles' heel of modern societies and states. Therefore, at a minimum, satellite-based communications require quantum-cryptography-capable software (like PQC) and, where feasible, hardware-enabled protections such as QKD. For that purpose, the development of new materials and components for quantum communication hardware that can withstand the space environment is also important.

---

[89] *See* e.g., Ria Chakraborty, Kim de Laat & Raymond Laflamme, *Canada's Migration to Post-Quantum Cryptography: Public-Private Roles*, CIGI Special Report (Oct. 29, 2025), https://www.cigionline.org/publications/canadas-migration-to-post-quantum-cryptography-public-private-roles/.

[90] Luxembourg Times, Feb. 4, 2026, Russian spy spacecraft have intercepted Europe's key satellites, officials believe, https://www.luxtimes.lu/world/russian-spy-spacecraft-have-intercepted-europes-key-satellites-officials-believe/128818976.html.

Anticipating this threat, the global community is developing two complementary responses. The first is PQC: new mathematical algorithms designed to withstand quantum attacks and to replace or supplement vulnerable classical cryptosystems. The second is quantum key distribution (QKD): protocols that use quantum physics to exchange encryption keys without relying on computational hardness assumptions.

The European Union has recognized quantum-secure communications as vital for safeguarding government data and critical infrastructure. In 2019, it launched the European Quantum Communication Infrastructure (EuroQCI) initiative, bringing together all 27 Member States and the European Space Agency (ESA) to build a continent-wide quantum-secure network. EuroQCI combines terrestrial fiber-QKD networks with a satellite component and has become a key part of the EU's cybersecurity strategy. The EU Quantum Technologies Flagship and OPENQKD have tested QKD in real-world conditions, and in 2022 the EU and ESA announced EAGLE-1, a quantum-communications satellite demonstrator planned for launch around 2025–26.[91]

QKD has advanced from basic physics experiments to field trials and early commercial use. Modern fiber-based QKD systems can operate at gigahertz repetition rates and deliver secure key rates of several kilobits per second over metropolitan distances of roughly 100 km using advanced protocols such as decoy-state BB84 or continuous-variable QKD. At the device level, single-photon detectors have improved markedly: superconducting nanowire single-photon detectors (SNSPDs) now exceed 90% detection efficiency at telecom wavelengths (1550 nm), with dark-count rates below 100 counts per second and timing jitter in the range of 50–150 ps. State-of-the-art WSi SNSPDs, for example, reach roughly 93% system detection efficiency at 1550 nm, significantly outperforming legacy InGaAs avalanche photodiodes, which typically offer only 20–25% efficiency and much higher dark counts. These advances in sources and detectors have enabled longer-distance QKD demonstrations, higher-rate exchanges, and multi-node QKD networks. China, notably, built a 2,000 km Beijing–Shanghai QKD backbone using trusted relays, while European efforts such as SECOQC and the Swiss Quantum Network have shown that networks connecting dozens of fiber-based QKD nodes are feasible.

Meanwhile, PQC has reached an important milestone through algorithm selection and standardization. In 2022, NIST announced the first set of quantum-resistant cryptographic standards, including CRYSTALS-Kyber for key exchange and CRYSTALS-Dilithium, FALCON, and SPHINCS+ for digital signatures.[92] Governments and companies have already begun testing these algorithms in conventional security protocols, including PQC-based VPNs,

---

[91] SES, EAGLE-1: Advancing Europe's Leadership in Quantum Communications, https://www.ses.com/newsroom/eagle-1-advancing-europes-leadership-quantum-communications.

[92] NIST, NIST Releases First 3 Finalized Post-Quantum Encryption Standards, Aug. 13 (2024), https://www.nist.gov/news-events/news/2024/08/nist-releases-first-3-finalized-post-quantum-encryption-standards.

PQC-TLS handshakes, and secure messaging systems. Early integration of QKD and PQC is also emerging in practice: several "quantum-safe" network pilots, including elements of South Korea's government network, are deploying QKD for key distribution while simultaneously using PQC-based encryption on data channels, thereby layering protection against failure in either approach.

In the space domain, the Chinese Micius satellite, launched in 2016, marked an early breakthrough for QKD by demonstrating quantum key exchange between a satellite and multiple ground stations, and even between distant ground stations via the satellite acting as a trusted relay over distances exceeding 1,000 km.[93] The secure key rates achieved were modest, but they established proof of concept for space-based QKD. More recent advances from Jian-Wei Pan's group, however, sharpen what it now means to be "ahead" in long-distance quantum networking beyond deployed QKD trunks. In 2026, the group reported long-lived memory-memory entanglement between remote trapped-ion nodes connected by 10 km of optical fiber, with coherence surviving longer than the average time required to establish the entanglement itself—directly addressing a central quantum-repeater bottleneck in which decoherence has historically outpaced entanglement generation and purification.[94] In a complementary breakthrough, the same broader research program demonstrated device-independent QKD at experimentally significant distances, including proof-of-principle finite-size device-independent QKD over 10 km and positive asymptotic key rates over 100 km with single atoms, pushing device-independent QKD toward city-scale regimes and reducing exposure to implementation side channels.[95] Although these remain laboratory demonstrations rather than fielded networks, and scaling to deployed fiber, field maintenance, and repeater-chained architectures will require sustained engineering, taken together they underscore China's momentum not only in backbone and satellite-scale quantum communications, but also in the repeater-grade entanglement primitives required for a future quantum internet. Upcoming missions such as EAGLE-1 in Europe aim to demonstrate QKD from low Earth orbit in a more operational setting, including attempts at daytime QKD through better background-light filtering, but the Pan-group results raise the comparative benchmark from network demonstration to durable, repeater-relevant quantum-network functionality.

A standards-first perspective is therefore not peripheral but operational. Interoperable benchmarks, certification pathways, crypto-agility requirements, and procurement rules can reduce fragmentation while allowing law and policy to adapt at the speed of technical change.[96] Recent public-private migration analysis likewise shows that successful PQC adoption depends not only on new algorithms but also on certification capacity, laboratory accreditation, common criteria, and sector-specific rollout discipline.[97]

[93] S. Liao et al., Satellite-to-ground quantum key distribution, Nature 549, 43 (2017), https://www.nature.com/articles/nature23655.
[94] Wen-Zhao Liu et al., Long-Lived Remote Ion-Ion Entanglement for Scalable Quantum Repeaters, Nature (2026), https://doi.org/10.1038/s41586-026-10177-4.
[95] Bo-Wei Lu et al., Device-Independent Quantum Key Distribution over 100 km with Single Atoms, 391 Science 592 (2026), https://doi.org/10.1126/science.aec6243.
[96] Aboy et al., *supra* note 3.

Despite these advances, no satellite QKD system to date has been designed for continuous, long-term operational service. Micius was fundamentally a physics experiment with limited duration and manually scheduled operations, and its payload components were not specifically hardened for multi-year in-orbit use. EAGLE-1 is similarly a three-year demonstrator. In short, the current state of the art shows that QKD is technically feasible over free-space links and that PQC is becoming deployable, but substantial gaps remain before either can be considered robust, integrated, and mission-ready for critical applications.

**Fragility to environment**

Quantum communication hardware, especially for satellites, is not yet fully designed to withstand extreme operational conditions. A single-photon source or detector that performs reliably in the laboratory may degrade or fail in space. Radiation can increase a detector's dark-count rate, quickly pushing the QKD quantum bit error rate (QBER) beyond acceptable limits and forcing the link to abort. Temperature fluctuations can misalign optics or gradually reduce device efficiency. Mechanical stresses such as launch vibration or micro-shocks in orbit can disturb sensitive optical alignments or detector settings. In operational terms, this means that a quantum-secure link may fail or degrade precisely when it is most needed.

**Fragility in continuity of security**

Current quantum communication systems often operate as overlays to classical networks rather than as fully integrated architectures. If a QKD link fails—because of a technical problem or simply because a satellite is out of view—systems often revert to classical key exchange methods that are not quantum-safe, such as RSA or ECDH. This creates security gaps. Strategic communications, especially military and diplomatic links, cannot tolerate periods of degraded security. More broadly, the absence of a unified architecture combining QKD with PQC obliges operators to maintain separate systems and policies, thereby reducing the effective value of QKD when operating conditions are suboptimal.

For allied deployment, this logic resembles a modern Bletchley method: tight feedback loops between science, engineering, operations, and alliance management; continuous testing and verification; and a certification compact that prevents PQC or QKD deployments from degrading into insecure fallback.[98] In policy terms, continuity of security requires crypto-agility by design, auditable migration playbooks, and trusted pathways for cross-border certification and incident response.

**Interoperability and scalability**

Most current QKD systems remain proprietary or confined to isolated pilot networks. Without interoperable standards, a QKD system from one vendor or country may not work with another, which complicates international or multi-party deployment. Scaling QKD into larger systems—with key management, repeaters, or multiple satellite hops—remains technically and institutionally difficult. PQC integration also requires protocol changes, accommodation of larger key sizes, and revised trust assumptions. This fragmentation hinders the growth of quantum-secure communications and underscores the need for a unified framework that can manage both cross-platform interoperability and the transition from classical encryption to quantum-safe security.

[97] Mauritz Kop, *A Bletchley Park for the Quantum Age*, War on the Rocks (Nov. 6, 2025), https://warontherocks.com/2025/11/a-bletchley-park-for-the-quantum-age/.
[98] *Ibid.*

Without intervention, these vulnerabilities may discourage investment in quantum security technologies or lead to only narrow, partial deployments in which the overall security gain remains modest.

### 5.2. Mission Assurance Beyond Communications

The logic developed here extends beyond communications. In mission settings, the relevant question is not only whether a quantum component performs under ideal conditions, but whether the system as a whole remains trustworthy under stress. That requires linking materials criticality, environmental fragility, and operational fallback behavior to concrete security outcomes. In practice, the same framework applies across quantum-secure communications, QS, and future quantum-enabled computing systems: components must either continue to operate securely, degrade predictably and safely, or trigger controlled suspension rather than silent failure. This mission-assurance perspective is especially important where systems are deployed in space, Arctic, defense, or other harsh environments in which repair, replacement, or recalibration may be delayed or impossible.

Normatively, this also points toward a governance stance of security-sufficient openness rather than either unqualified openness or indiscriminate closure.[99] Mission assurance depends on preserving enough openness for standards, benchmarking, trusted allied coordination, and legitimate scientific exchange, while reserving targeted controls for genuinely sensitive capabilities and chokepoints.

### 5.3. Quantum-AI Hybrids: AI & ML Convergence with QC

Quantum-AI hybrids create a distinct security surface because they couple hardware-constrained quantum workflows—often remote, cloud-mediated, and vendor-specific—with data- and model-centric AI pipelines. In practice, near-term convergence is most likely to occur through (i) quantum subroutines embedded in classical machine-learning training or inference loops, (ii) quantum-enhanced sampling and optimization workflows, and (iii) AI-assisted compilation, calibration, and error mitigation for quantum processors. This coupling shifts the security question from “Is the quantum device secure?” to “Is the end-to-end socio-technical pipeline secure across data, models, runtimes, and hardware dependencies?”

[99] Kop, Nexus, *supra* note 17.

From a mission-assurance perspective, hybrid systems must preserve confidentiality, integrity, and availability under both cryptographic transition risk—PQC migration and any QKD dependencies—and model-security failure modes, including data poisoning, model inversion or extraction, supply-chain compromise of libraries and control software, and insecure fallback paths. Criticality should therefore be assessed not only for qubit materials and cryogenic toolchains, but also for enabling compute such as GPUs and accelerators, trusted execution environments, orchestration layers, and secure update mechanisms. Compromise at any of these layers can negate the security benefits expected from quantum-secure communications or quantum-enabled analytics.

U.S.–China strategic competition intensifies these risks by shaping where hybrid workloads run, which vendors are trusted, and how export controls and procurement rules affect access to advanced compute, quantum hardware, and the CMMs that underpin both. Quantum-AI convergence should therefore be treated as a cross-domain supply-chain and deployment-security problem in which continuity of security depends on robust governance of the interfaces among data, models, runtimes, and quantum hardware rather than on any single technical component.

In addition to using AI to help develop quantum computers, research on hybrid QC is actively underway. This field seeks to leverage the computational advantages of quantum systems where classical algorithms, including conventional AI methods, encounter structural limits. While data-driven machine learning and deep learning have dramatically expanded the range of solvable problems, they still rely in practice on approximation, extensive training, and hyperparameter tuning because perfect global optimization is often infeasible. Current AI systems are therefore best understood as powerful but bounded approximators.[100]

Quantum computers, by contrast, are attractive because they may accelerate certain classes of highly complex computations involving probabilistic distributions, optimization, or quantum-system simulation. This has led to growing interest in QC as a possible resource for next-generation AI. At present, however, the connection between QC and AI has not yet reached mature commercial deployment. Rather, hybrid QC is evolving into a quantum-HPC paradigm that combines AI, classical high-performance computing, and quantum processors, assigning classical systems the tasks they perform efficiently while reserving quantum systems for computations where quantum advantage may eventually emerge.

To enable this, a preprocessing step is often required in which classical data are encoded into forms suitable for qubits, and quantum circuits are initialized in appropriate states. One common approach uses a quantum feature map (QFM), a mapping function that converts classical data into quantum states. By mapping classical data into Hilbert space, a high-dimensional quantum state space, it may become possible to identify complex structures and patterns not visible in low-dimensional representations. The design of the quantum feature map is therefore important, and AI may help optimize the mapping structure and parameters through learning.

[100] J.M. Kim, SPRi Issue Report IS-204, Convergence of quantum computing and AI: Potential advances and implications, Software Policy & Research Institute (July 21, 2025).

The quantum feature map transforms classical data into quantum state space, while the quantum kernel expresses a similarity function between transformed quantum states. By mapping classical data into quantum state space and projecting it back into classical space, hidden structures may become easier to identify.[101] For example, as shown in figure 8 below, complex nonlinear decision boundaries that are difficult to distinguish in two-dimensional input space may become easier to separate once mapped into higher-dimensional space, such as 3D vector space.[102]

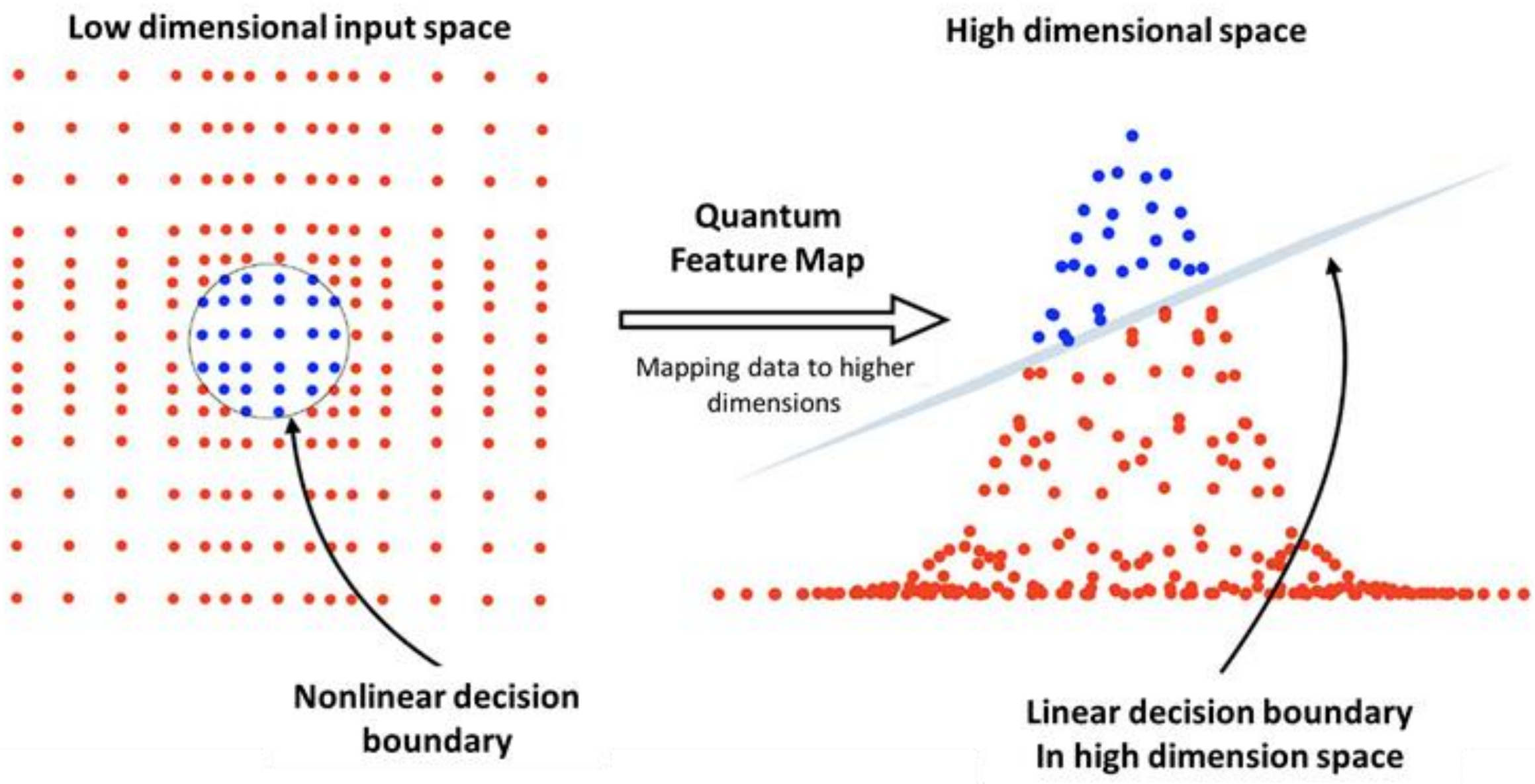

*Figure 8. Classical data encoding using quantum feature maps.*[103]

Quantum AI (QAI) refers to the broader research field that seeks to use the potential computational advantages of quantum computers to develop AI methods such as machine learning, neural networks, and large language models. Interest in QAI reflects both significant advances in quantum hardware and the limits of the computing resources required for large-scale AI training and inference. Representative areas include quantum machine learning (QML), quantum neural networks (QNN), and quantum optimization (QO).[104]

---

[101] Huang, H. Y. et al., Power of data in quantum machine learning. Nature communications, 12(1), 2631 (2021)
[102] Dataiku. (March 20, 2024). Quantum leap: Beyond the limits of machine learning. Dataiku Blog, https://blog.dataiku.com/quantum-leap-beyond-the-limits-of-machine-learning; https://medium.com/data-from-the-trenches/quantum-leap-beyond-the-limits-of-machine-learning-af7f3c292b75
[103] *Ibid.*
[104] Wang, X. et al., Quantum Artificial Intelligence for Software Engineering: the Road Ahead. ACM Trans. Softw. Eng. Methodol., Vol. 37, No. 4, Article 111. (2025).

Quantum machine learning, a core subfield of QAI, generally focuses on accelerating machine-learning tasks using quantum computers. This work may be grouped into (i) using classical machine learning to perform quantum-related tasks, (ii) using quantum computers to accelerate classical machine learning, and (iii) learning quantum data using quantum computers.[105] Quantum machine-learning algorithms can also be classified into supervised, unsupervised, and reinforcement-learning settings. Commonly discussed approaches include quantum support vector machines, quantum neural networks, quantum k-means clustering, and quantum PCA.[106]

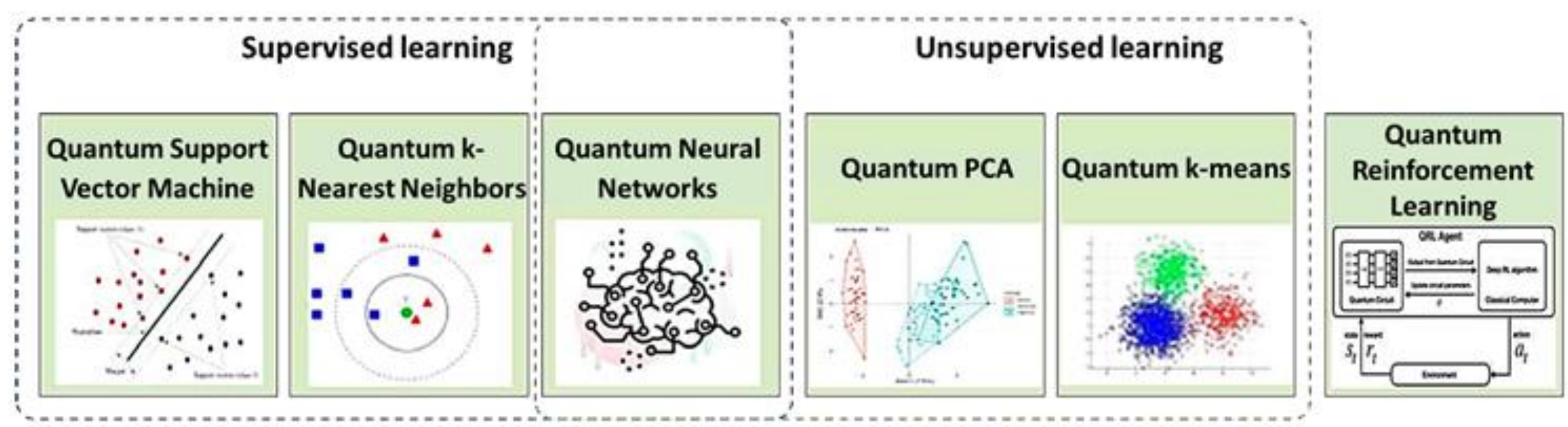

*Figure 9. Examples of Quantum Machine Learning Algorithms.*[107]

Quantum machine learning uses quantum kernels to map data into high-dimensional quantum state spaces and may thereby help reveal complex structures in classification and generative tasks. For example, IonQ has reported that, for a specific generative modeling task, a trapped-ion quantum computer-based model produced a result distribution closer to the target distribution after 28 iterations than a classical GAN after 20,000 iterations. While such findings should be interpreted cautiously and do not establish broad commercial superiority, they illustrate why quantum-enhanced AI continues to attract attention.[108]

[105] Beer, K et al., Training deep quantum neural networks. Nature communications, 11(1), 808 (2020).
[106] Chen, L. et al., Design and analysis of quantum machine learning: a survey. Connection Science, 36(1), 2312121 (2024).
[107] *Ibid.*
[108] IonQ IR presentation (May 8, 2024), https://s28.q4cdn.com/828571518/files/doc_financials/2024/q1/24-05-08-Investor-Updates-Q1-2024.pdf, Generative Quantum Machine Learning for Finance (Jan 8, 2025), https://www.ionq.com/resources/generative-quantum-machine-learning-for-finance.

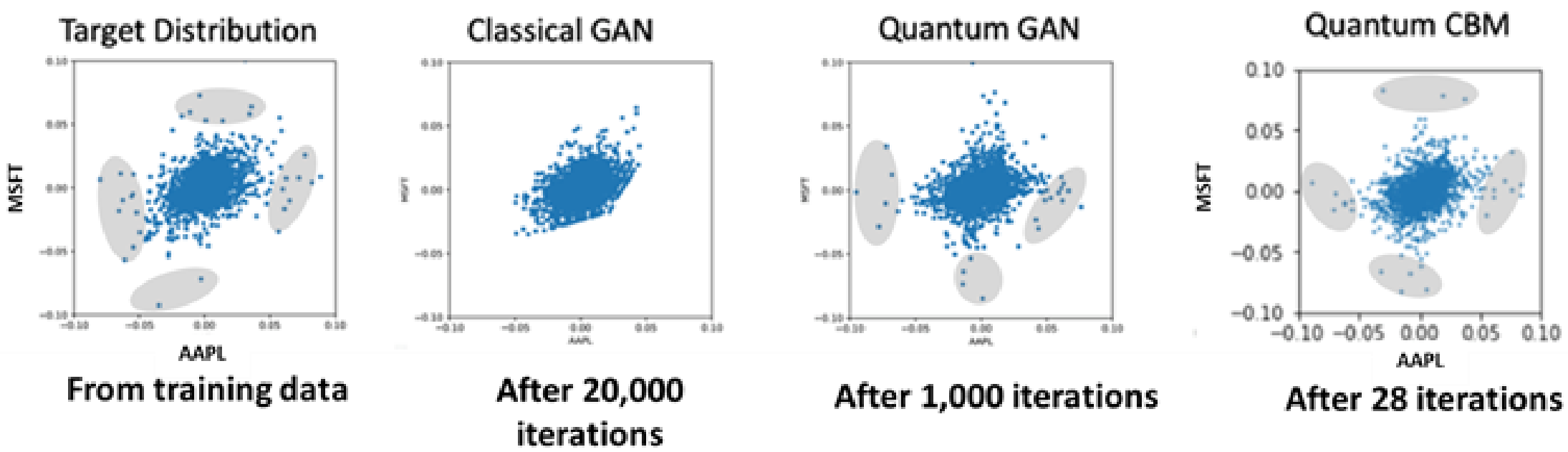

*Gray area: outliners resulting from notable events

*Figure 10. Examples of enhancing data learning effectiveness using quantum computers.*[109]

The emergence of AI and the recent acceleration of QC have together marked an important inflection point. AI is already being used in the stack to help identify and correct quantum errors and to improve solutions to existing technical bottlenecks. These developments are often grouped into two related directions: "AI for Quantum," meaning the use of AI to advance QC, and "Quantum for AI," meaning the use of QC to accelerate AI. As interest in new quantum materials and hardware grows, so too does interest in their practical use as next-generation computing resources for AI development. The growing use of AI in quantum computer development is thus accelerating the broader evolution of a quantum-classical hybrid computing environment.

AI-quantum convergence also compresses governance timelines because AI already supports calibration, control, error mitigation, and elements of materials discovery and design. As a result, supply-chain governance, model governance, and cloud governance are becoming increasingly entangled.[110] The policy consequence is that continuity of security will depend less on any single device and more on disciplined governance of interfaces among data, models, standards, control software, and quantum hardware.

### 5.4. QC, QS, and Networking in LiDAR Gravitational Detection and Technological Surprise

QS and photonic networking are also becoming relevant to strategic sensing and detection infrastructures. A recent example is a compact, lightweight single-photon airborne LiDAR system capable of acquiring high-resolution 3D images with a low-power laser, potentially making single-photon LiDAR practical for air and space applications such as environmental monitoring, 3D terrain mapping, and object identification.[111] The system works by sending laser pulses toward the ground and capturing the reflected photons with highly sensitive single-photon avalanche diode (SPAD) arrays. Computational imaging algorithms then reconstruct high-resolution 3D terrain images, including daytime imaging over large areas (230×470$m^2$; 754×1542 $ft^2$).

[109] *Ibid.*
[110] Cho, Kop & Lee, *supra* note 5.
[111] Y. Hong et al., Airborne single-photon LiDAR towards a small-sized and low-power payload, Optica, 11 (5), 612 (2024), https://doi.org/10.1364/OPTICA.518999.

The single-photon LiDAR arrays are synthesized from three InGaAs/InP 64×64 SPAD arrays, corresponding to 12,288 photonic quantum sensor chips. Such systems enable highly accurate 3D mapping even in challenging environments such as dense vegetation or urban settings. They are also attractive for deployment on resource-constrained drones and satellites, which requires shrinking the overall system and reducing energy consumption. As one recent example shows, Chinese scientists have developed advanced airborne LiDAR systems that use exceptionally low laser power and a small optical aperture while still achieving good detection range and imaging resolution through sub-pixel scanning and a new 3D deconvolution algorithm. Ground testing showed lidar imaging with a resolution of ≤6 cm from 1.5 km away.

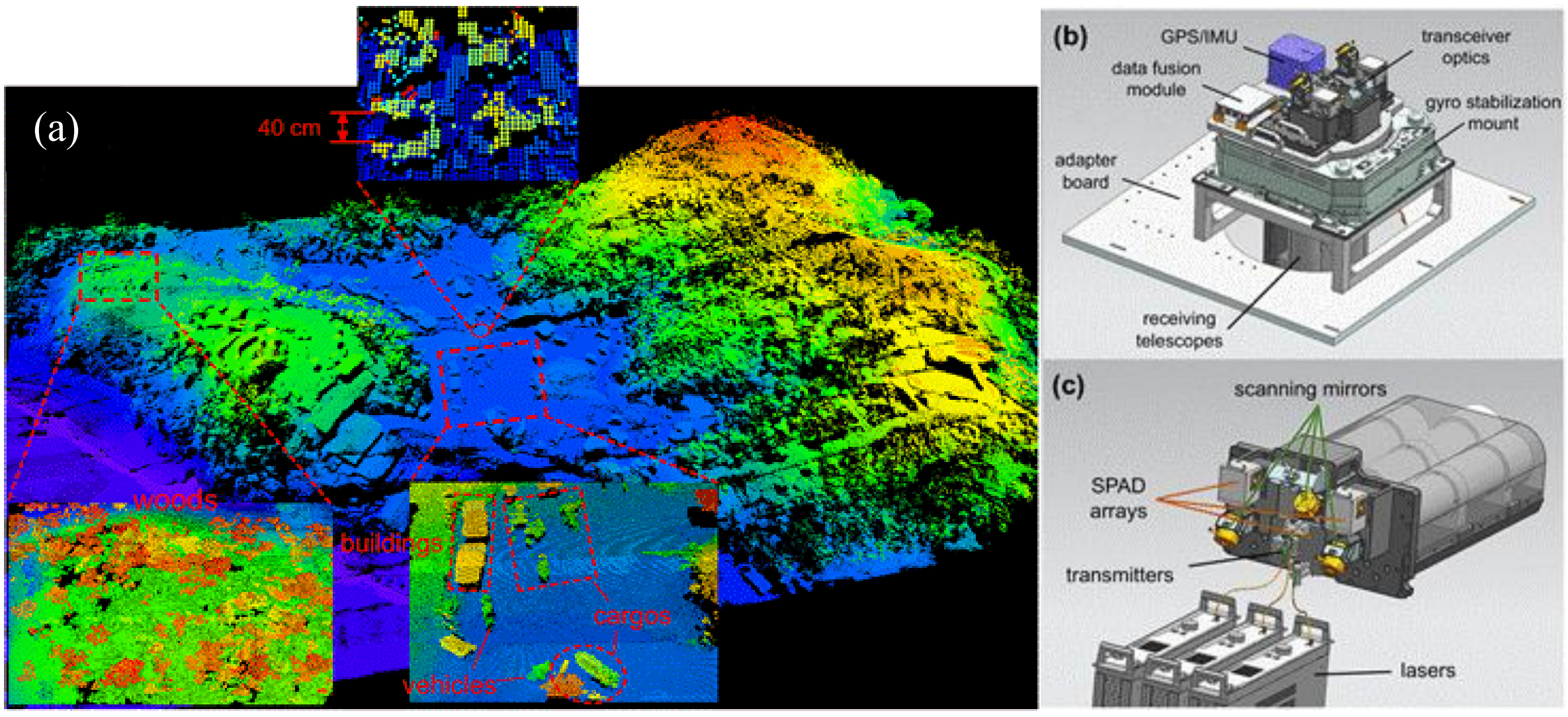

*Figure 11. (a) Example 3D imaging result of a 230m×470m region. The color encodes the height and different color maps are used for the sub-figures and the main figure (b) Main components of the system installed on the gyro stabilization mount. (c) Details of the transceiver optics.*[112]

At the same time, the growing demand for quantum chips in scaling quantum computers, quantum sensors, and quantum communications is driving the rapid build-out of quantum foundries. Foundries capable of manufacturing barium titanate and thin-film lithium niobate (TFLN) wafers—materials widely used not only in photonic quantum computers but also in quantum sensors and quantum communication chips—are now being launched in rapid succession.

In May 2025, PsiQuantum reported a 300-mm barium titanate (BTO) wafer manufacturing platform — integrating single-photon sources, superconducting single-photon detectors, and quantum benchmarking circuit components within a fully integrated silicon photonics process

[112] *Ibid.*

flow — fabricated at GlobalFoundries as a commercial semiconductor foundry (see Table 6).[113] Also in May 2025, a 150 mm thin-film lithium niobate (TFLN) wafer manufacturing platform was launched as a quantum photonic chip foundry intended to accelerate the commercialization of photonic QC. The platform optimized TFLN etch processes to 0.3 nm sidewall roughness and targets scalable production of TFLN devices, including electro-optic modulators, phase modulators, ring resonators, filters, nonlinear PPLN waveguides, and micro-rings, to meet growing customer demand in emerging quantum-computing markets.[114]

China is moving quickly as well. In June 2025, Shanghai Jiao Tong University's Chip Hub for Integrated Photonics Xplore (CHIPX) announced completion of a pilot manufacturing line capable of producing 6-inch wafers made from thin-film lithium niobate, a material increasingly seen as critical to high-performance photonics. The new facility is China's first photonic quantum-chip manufacturing platform and has an annual capacity of 12,000 6-inch wafers with rapid, low-cost production.[115] Assuming roughly 350 integrated chips per 6-inch wafer, the CHIPX facility would have a production capacity of approximately 4.2 million photonic chips per year, sufficient to support growing future demand.

These quantum-chip foundries enable real-time testing on automated mass production lines, replacing the traditional manual, one-by-one manufacturing and testing process typical of laboratory settings. This compresses manufacturing timelines from several months to one or two weeks and materially improves prospects for rapid commercialization and scaling. These quantum-chip foundries generally and essentially require molecular beam epitaxy (MBE) systems to produce TFLN and BTO wafers as well as III–V compound semiconductor materials.[116] As is often the case in fragile supply chains, only a small number of companies specialize in the commercial manufacture of MBE systems. The leading suppliers are RIBER (France) and Veeco (United States), while other significant players include SVT Associates (United States), DCA Instruments (Finland), CreaTec Fischer & Co. GmbH (Germany), and AdNanoTek (Taiwan).[117] The reason for configuring the mass production line in China with 6-inch wafers, which are different from Western 8-inch or 300 mm wafers, is that equipment for manufacturing and producing wafers larger than 6 inches are not easy to include their supply chain to date. More broadly, these developments matter for the present paper because they show how materials availability, foundry capacity, and advanced sensing architectures can together create pathways for technological surprise in both civil and security domains, including remote sensing, navigation, secure communications, and future gravitational or space-based detection infrastructures.

---

[113] PsiQuantum team, A manufacturing platform for photonic quantum computing, Nature 641, 876 (2025), https://doi.org/10.1038/s41586-025-08820-7.
[114] Arizona Commerce Authority, Quantum computing Inc. celebrates grand opening of quantum photonic chip foundry in Tempe (May 12, 2025), https://www.azcommerce.com/news-events/news/2025/5/quantum-computing-inc-celebrates-grand-opening-of-quantum-photonic-chip-foundry-in-tempe/, https://quantumcomputinginc.com/foundry.
[115] China Ramps Up Photonic Chip Production with eye on AI and Quantum Computing, Quantum Insider (June 2025), https://thequantuminsider.com/2025/06/13/china-ramps-up-photonic-chip-production-with-eye-on-ai-and-quantum-computing/.
[116] *See*: A. L. Pellegrino et al., Adv. Mater. Interfaces, 10 (34), 2300535 (2023), Efficient Optimization of High-Quality Epitaxial Lithium Niobate Thin Films by Chemical Beam Vapor Deposition: Impact of Cationic Stoichiometry, https://doi.org/10.1002/admi.202300535; and W. Nunn et al., C. Grivas, J. Vac. Sci. Technol. A 39, 040404 Hybrid molecular beam epitaxy growth of BaTiO3 films (2021), https://doi.org/10.1116/6.0001140.
[117] RIBER S.A., *Molecular Beam Epitaxy (MBE) Products and Services*, https://www.riber.com (last visited Apr. 20, 2026).

| Year | Quantum chip wafer | | | |
|---|---|---|---|---|
| | US | | China | |
| | Player | Type | Player | Type |
| | Manufacturing | Photon | Manufacturing | Photon |
| Feb. 2025 | PsiQuantum + Global Foundries | BTO 300mm wafer | | |
| May 2025 | Quantum computing Inc | TFLN 150mm wafer | | |
| June 2025 | | | CHIPX TFLN | 6-inch wafer 12,000/yr* |

*Table 6. Quantum Chip Wafer's manufacturing (*annual production capacity).*[118]

### 6. Conclusion: Towards Geostrategic Critical Minerals and Materials Resilience

This analysis suggests that quantum supply-chain risk is best evaluated through a sector-specific criticality lens that links upstream concentration and policy volatility to downstream performance requirements. The two use cases—niobium and nickel dependence in superconducting platforms and the space deployment of SNSPDs or LiDAR—illustrate that vulnerability is shaped not only by raw-material availability, but also by refining chokepoints, qualification timelines, and system-level requirements for components to fail gracefully under harsh operating conditions. For the scaling and manufacturing of QT, the U.S. chokepoint profile is concentrated in CMM and processing technologies, with human capital shifting from a non-critical constraint at present to a potentially near-critical one in the future. By contrast, China's profile reflects a near-critical human-capital constraint in the present that may ease over time, while financial investment and raw materials appear comparatively less constraining in both the present and near future.

More specifically, the paper argues that geostrategic resilience in QT should be built on four mutually reinforcing layers: dynamic criticality assessment rather than static lists; standards-and-certification pathways that preserve interoperability and auditability; stockpiling and diversification strategies focused on processed, qualified, and deployable inputs rather than raw tonnage alone; and mission-assurance design principles that treat environmental survivability and continuity of security as first-order requirements. To operationalize this framework, we propose the development of a dedicated Quantum Criticality and Critical Minerals (QCCM) dashboard: a living, continuously updated decision-support tool that makes visible changes in material concentration, refining dependence, qualification bottlenecks, stockpiling gaps, substitutability, and geopolitical stress signals across mission-relevant quantum platforms. Such a dashboard would translate abstract criticality analysis into a practical early-warning and prioritization instrument for governments, allied partners, and industry. Translated into security and deployment terms, quantum–AI hybrids offer a near-term pathway for mission assurance by combining presently immature quantum subsystems with classical AI and high-performance computing, with quantum-secure communications as one anchoring use case. The bismuth episode analyzed in Section 1 and depicted in Figure 3 illustrates the QCCM dashboard's intended operational value: a licensing action in one jurisdiction produced a tenfold spot-price move within thirty days, well inside the revision cycle of any static national criticality list.

[118] Min-Ha Lee, *Critical Materials for Next Generation: Quantum Information Technologies*, (Critical Materials Innovation Hub US DOE annual meeting, Idaho National Laboratory, Aug. 28, 2025), *supra* note 14.

In this sense, the manuscript advances a geostrategic critical minerals and materials resilience agenda: one that centers secure supply-chain analysis and criticality assessment for quantum technologies deployed in Arctic, aerospace, and space environments, and that defines resilience as the capacity to scale, field, and sustain mission-relevant QT capabilities despite concentrated upstream dependencies, operational fragility, and rapidly shifting policy conditions.

These technical dependencies are now embedded within a broader context of U.S.–China strategic competition over rare earths and adjacent critical materials. China has recently advanced an international partnership track through the International Economic and Trade Cooperation Initiative on Green Mining and Minerals[119], while the United States and its allies are simultaneously pursuing diversification agreements, domestic capacity-building, and alternative supply-chain partnerships. Against this backdrop, the proposed indicator set and mission-stressor-to-metric mapping offer a reproducible starting point for prioritizing mitigations, including design-around, diversification, stockpiling, shielding, and qualification strategies.

Future supply chain criticality work should expand the evidentiary base through platform-specific bills of materials and empirically grounded substitution and qualification pathways, thereby enabling more robust comparison across architectures, sectors, and mission profiles. Taken together, the strategic question is no longer whether quantum technologies will matter, but whether states and allied ecosystems can secure the materials, standards, and mission-assurance architectures needed to ensure that quantum capability is not merely powerful, but trustworthy, resilient, and governable—equal to the geostrategic pressures that will define the quantum age.

<center>* * *</center>

[119] State Council of China, China welcomes more countries, int'l organizations to join green mining, minerals cooperation initiative: ministry, Nov 27, 2025, https://english.www.gov.cn/news/202511/27/content_WS69284a29c6d00ca5f9a07d2c.html.